\documentclass{iopart}

\usepackage{epsfig}
\usepackage{graphicx}
\usepackage{hhline}

\begin{document}

\title{Study of nucleon resonances with electromagnetic interactions}

\author{(From EBAC, Thomas Jefferson National Accelerator Facility)\footnote{Notice: Authored by Jeffers
on Science Associates, LLC under U.S. DOE 
Contract No. DE-AC05-06OR23177. The U.S. Government retains a 
non-exclusive, paid-up, irrevocable, world-wide license to publish or 
reproduce this manuscript for U.S. Government purposes.}}

\author{T.-S. H. Lee$^1$, L.C. Smith$^2$}

\address{$^1$Physics Division, Argonne National Laboratory, Argonne,
 Illinois 60439, USA \\
and \\Excited Baryon Analysis Center,
Thomas Jefferson National
Accelerator Facility, Newport News, Virginia 23606, USA}
\ead{lee@phy.anl.gov}

\address{$^2$Physics Department, University of Virginia, Charlottesville, Virginia 22901, USA}

\begin{abstract}
Recent developments in using electromagnetic meson production reactions
to study the structure of nucleon resonances are reviewed.
Possible future works are discussed. 
\end{abstract}

\section{Introduction}
The study of nucleon resonances ($N^*$) has long been recognized as
an essential step towards developing a fundamental understanding of strong
interactions.  Since the discovery of the $\Delta$(1232) in 1951, many 
$N^*$'s have been identified mainly through amplitude analyses of 
pion-nucleon elastic scattering. This effort peaked in the 1970's 
with fairly consistent results from several independent groups. 
A long-standing challenge in hadronic physics is to understand 
the spectroscopy of these resonant states, including their 
electromagnetic and strong
decays, within a framework consistent with QCD. However, many of the 
assumptions used in phenomenological analyses remain to be justified 
theoretically. Also, the limitations of the data base lead to rather
large errors in many of the determined resonance parameters.

Some additional but limited pion-nucleon scattering data have 
been obtained since the 1980's. While these data have improved the 
determination of $\pi N$ amplitudes at low energies, there has not
been much impact in the $N^*$ region.  The situation changed 
drastically in the 1990's with the construction of new electron and 
photon beam facilities.  Experiments at Jefferson Laboratory, MIT-Bates, 
LEGS, Mainz, Bonn, GRAAL, and Spring-8 are providing new data on the 
photo- and electroproduction of $\pi$, $\eta$, $K$, $\omega$, $\phi$ mesons 
and $2\pi$ final states.  These data offer a new opportunity to
understand in detail how $N^*$ properties emerge from the non-perturbative 
aspects of QCD. For example, the extraction of $\gamma N\rightarrow N^*$ transition form factors could shed light on the dynamical origins of the confinement of constituent quarks and the associated meson cloud. 

To make progress, we need to perform amplitude analyses of the data 
to extract the $N^*$ parameters. More importantly,
we  need to develop dynamical reaction
models to interpret the extracted 
$N^*$ parameters in terms of QCD. At the present time,
the achievable goal is to test
the predictions from various QCD-based 
hadron models such as the well-developed constituent quark model\cite{capstick}
and the covariant model based on Dyson-Schwinger Equations\cite{roberts}. 
In the near future,
we hope to understand the $N^*$ parameters in terms of Lattice QCD (LQCD).

In the $\Delta$ region, both the amplitude analyses and reaction models
have been well developed.
We find that these two efforts are complementary.
For example, a recent review of amplitude analyses~\cite{arndt} gave the 
result that the $N$-$\Delta$ M1 transition strength is  $G_M(0)=3.18\pm0.04$.
This value is about 40 $\%$ larger than the constituent quark model predictions 
$G_M(0)=\frac{4}{3}\sqrt{\frac{m_N}{2m_\Delta}}G_p(0)\sim 2.30$.
Using dynamical models~\cite{sl,dmt} one can show that the discrepancy is 
due to the pion cloud which is not included in the constituent quark model 
calculations.  The effect of the meson cloud on the $N$-$\Delta$ transition
has been further revealed in the study of pion electroproduction reactions.

In the higher mass $N^*$ region, the situation
 is much more complicated because of many open channels.
Any reliable analysis of the meson production data in this 
higher energy region must be
based on a coupled-channel approach. In recent years, significant progress
has been made in this direction.

In this article, we will briefly summarize the status of theoretical
and experimental efforts in the study of $N^*$ resonances.
In section 2, we review most of the current meson production reaction models
used to extract $N^*$ information from meson production data.
In section 3, we summarize the results from recent analyses of photo- and 
electro-production data.  The focus will be on the $N-\Delta$ (1232) transition 
form factors which have now been investigated in great detail. We will also 
summarize the $N$-$N^*$ transition form factors extracted recently
by several groups. Discussions on future developments will be given in
section 4.

\section{Theoretical developments}

\begin{figure}[t]
\centerline{\includegraphics[width=12.0cm]{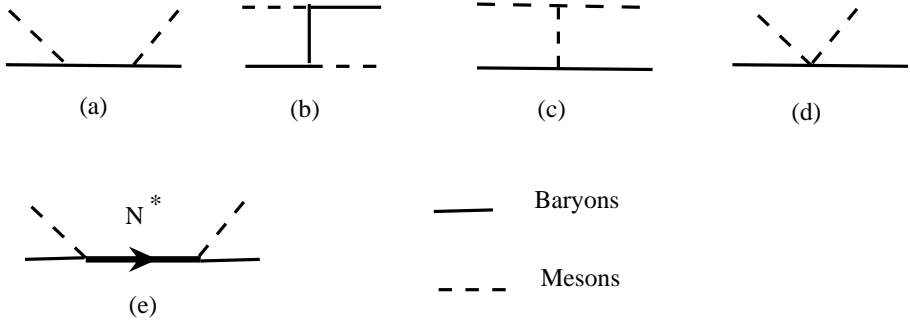}}
\caption[]{ Tree-diagram mechanisms of meson-baryon interactions.}
\label{fig:mechanism_1}
\end{figure}

Most of the models for meson production reactions can be derived by
considering the
following coupled-channel equations
\begin{eqnarray}
T_{a,b}(E) = V_{a,b} + \sum_{c}V_{a,c} g_c(E) T_{c,b}(E) \,,
\label{eq:tmatrix}
\end{eqnarray}
where
$a,b,c$  = $\gamma N$, $\pi N$, $\eta N$, $\omega N$, $KY$,
$\pi\Delta$, $\rho N$ $\sigma N$, $\cdot\cdot$. The interaction term is defined by
$V_{a,b} =<a|V|b>$ with
\begin{eqnarray}
V = v^{bg} + v^R \,.
\label{eq:v}
\end{eqnarray}
Here $v^{bg}$ represents the non-resonant(background) mechanisms such
as the tree diagrams illustrated in
 Figs.~\ref{fig:mechanism_1}(a)-(d), and $v^R$ describes
 the $N^*$ excitation shown in Figure~\ref{fig:mechanism_1}(e).
Schematically, the resonant term can be written as
\begin{eqnarray}
v^R(E) = \sum_{N^*_i}\frac{\Gamma^\dagger_i \Gamma_i}{E-M^{0}_i}\,,
\label{eq:vr}
\end{eqnarray}
where $\Gamma_i$ defines the decay of the $i$-th $N^*$ state into meson-baryon
states, and $M^0_i$ is a mass parameter related to the resonance position.

The meson-baryon propagator in (\ref{eq:tmatrix}) is
\begin{eqnarray}
g_c(E) = < c \mid g(E) \mid c > \,, \nonumber
\end{eqnarray}
with
\begin{eqnarray}
g(E)&=&\frac{1}{E-H_0 +i\epsilon} \, \nonumber \\
&=& g^P(E) -i\pi\delta(E-H_0)  \,,
\label{eq:t-prop}
\end{eqnarray}
where $H_0$ is the free Hamiltonian and
\begin{eqnarray}
g^P(E) &=& \frac{P}{E-H_0} \,.
\label{eq:k-prop}
\end{eqnarray}
Here $P$ denotes taking the principal-value part of any
integration over the propagator.
If $g(E)$ is replaced by $g^P(E)$ and $T_{a,b}(E)$ by $K_{a,b}(E)$,  
(\ref{eq:tmatrix}) then defines
the K-matrix which is related to T-matrix by
\begin{eqnarray}
T_{a,b}(E)= K_{a,b}(E) - \sum_{c}T_{a,c}(E)[i\pi\delta(E-H_0)]_c K_{c,b}(E) \,.
\label{eq:t-k}
\end{eqnarray}

By using the two potential formulation,
one can cast (\ref{eq:tmatrix}) into
\begin{eqnarray}
T_{a,b}(E)  &=&  t^{bg}_{a,b}(E) + t^{R}_{a,b}(E)  
\label{eq:t-2pot}
\end{eqnarray}
with 
\begin{eqnarray}
t^{R}_{a,b}(E) &=& \sum_{N^*_i, N^*_j}
\bar{\Gamma}^\dagger_{N^*_i, a}(E) [G(E)]_{i,j}
\bar{\Gamma}_{N^*_j, b}(E)  \,.
\label{eq:t-res}
\end{eqnarray}
The first term of (\ref{eq:t-2pot}) 
 is determined only by the non-resonant interaction 
\begin{eqnarray}
t^{bg}_{a,b}(E)= v^{bg}_{a,b} +\sum_{c} v^{bg}_{a,c}g_c(E) t^{bg}_{c,b}(E)\,.
\label{eq:t-nonr}
\end{eqnarray}
The resonant amplitude (\ref{eq:t-res}) is determined by the
dressed vertex
\begin{eqnarray}
\bar{\Gamma}_{N^*,a}(E)  &=&
  { \Gamma_{N^*,a}} + \sum_{b} \Gamma_{N^*,b}
g_{b}(E)t^{bg}_{b,a}(E)\,,
\label{eq:dressed-f}
\end{eqnarray}
and the dressed propagator
\begin{eqnarray}
[G(E)^{-1}]_{i,j}(E) = (E - M^0_{i})\delta_{i,j} - \Sigma_{i,j}(E) \,.
\label{eq:nstar-prop}
\end{eqnarray}
Here the mass-shift term is 
\begin{eqnarray}
\Sigma_{i,j}(E)= \sum_{a}\Gamma^\dagger_{N^*,a} g_{a}(E)
\bar{\Gamma}_{N^*_j,a}(E)\,.
\label{eq:nstar-sigma}
\end{eqnarray}
Note that the meson-baryon propagator $g_a(E)$ 
for channels including an unstable particle, such as
$\pi \Delta$, $\rho N$ and $\sigma N$,
must be modified
to include a width.

With the above equations, we now can derive in the next few subsections
most of the recent models for analyzing the data of
meson production reactions.

\subsection{Unitary Isobar Model}
In recent years, the most widely used model is 
the Unitary Isobar Model (UIM) developed~\cite{maid} by the Mainz group.
This model, called MAID,  
is based on the on-shell relation (\ref{eq:t-k}).
By including only one hadronic channel, $\pi N$ (or $\eta N$ ), 
(\ref{eq:t-k}) leads to
\begin{eqnarray}
T_{\pi N,\gamma N} = 
e^{i\delta_{\pi N}}cos\delta_{\pi N} K_{\pi N, \gamma N}\,,
\label{eq:uim-1}
\end{eqnarray}
where $\delta_{\pi N}$ is the pion-nucleon scattering phase shift.
By further assuming that 
$K \rightarrow V=v^{bg} + v^R$, (\ref{eq:uim-1}) can be cast into
the following form
\begin{eqnarray}
T_{\pi N,\gamma N}({\it UIM}) =
e^{\delta_{\pi N}}cos\delta_{\pi N} [v^{bg}_{\pi N, \gamma N}]
+\sum_{N^*_i}T^{N^*_i}_{\pi N, \gamma N}(E) \,.
\label{eq:uim-2}
\end{eqnarray}

The non-resonant term $v^{bg}$ in (\ref{eq:uim-2})  
is calculated from the standard Born terms but with an
 energy-dependent mixture of
pseudo-vector (PV) and pseudo-scalar (PS) $\pi NN$ coupling.
For the resonant term in (\ref{eq:uim-2}), MAID
uses the following parameterization by Walker~\cite{walk}
\begin{eqnarray}
T^{N^*_i}_{\pi N, \gamma N}(E) =f^i_{\pi N}(E)
\frac{\Gamma_{tot}M_i e^{i\Phi}}{M^2_i-E^2 - i M_i\Gamma^{tot}}
f^i_{\gamma N}(E)\bar{A}^i \,,
\label{eq:uim-3}
\end{eqnarray}
where $f^i_{\pi N}(E)$ and $f^i_{\gamma N}(E)$ are the form factors describing
the decays of $N^*$, $\Gamma^{tot}$ is the total decay width, and 
$\bar{A}^i$ is the $\gamma N \rightarrow N^*$ excitation strength.
The phase $\Phi$ is required by the unitarity condition and is determined by
 an assumption  that relates the phase of the total 
photoproduction amplitude to  the 
$\pi N$ scattering phase shift.

The UIM developed by the JLab-Yerevan collaboration~\cite{azn1,azn2,azn3} is similar to MAID.
The main difference is that this model uses the
Regge parameterization to define the 
amplitudes at high energies. Schematically, they write non-resonant amplitude
$[Background] = [e^{\delta_{\pi N}}cos \delta_{\pi N}v^{bg}]$ of 
(\ref{eq:uim-2}) as
\begin{eqnarray}
[Background] &=& [Born + \rho + \omega]_{UIM} \,\,\,\, at \,\,\ s < s_0 \\
&=& [Born + \rho + \omega]_{UIM}\frac{1}{1+(s-s_0)^2} \nonumber \\
& & + [Regge-Poles]\frac{(s-s_0)^2}{1+(s-s_0)^2}\,\,\,\, at \,\, s > s_0
\end{eqnarray}
where $s_0\sim 1.2$~GeV was determined phenomenologically in fitting
the pion photoproduction data.

Both MAID and JLab-Yeveran UIM have been applied extensively to analyze the
data of $\pi$ and $\eta$ production reactions. 

\subsection{K-matrix Coupled-Channel Models}
\subsubsection{VPI-GWU Model}
The VPI-GWU~\cite{said} (SAID) can be derived from (\ref{eq:t-k})
 by considering three channels:
$\gamma N$, $\pi N$, and $\pi\Delta$.
The solution of the resulting 
 $3\times 3$ matrix equation leads to
\begin{eqnarray}
T_{\gamma N,\pi N}(SAID) = A_I(1 + iT_{\pi N,\pi N}) 
+ A_RT_{\pi N,\pi N}\,,
\label{eq:said}
\end{eqnarray}
where 
\begin{eqnarray}
A_I &=& K_{\gamma N, \pi N} 
- \frac{K_{\gamma N ,\pi\Delta}K_{\pi N,\pi N}}
{K_{\pi N,\pi\Delta}} \,, \\
A_R&=&\frac{K_{\gamma N, \pi\Delta}}{K_{\pi N, \pi\Delta}}\,.
\end{eqnarray}
The  $\pi N$ amplitude $T_{\pi N,\pi N}$ needed to
evaluate (\ref{eq:said}) is available in SAID.
In actual analyses, they simply parameterize $A_I$ and $A_R$ as
polynomial functions of the on-shell momenta for pion and photon with
adjustable parameters.
Once the parameters of $A_I$ and $A_R$
have been  determined by fitting the pion photoproduction data,
the $N^*$ parameters are extracted by
fitting the resulting amplitude $T_{\gamma N,\pi N}$ at energies near the
resonance position to 
a Breit-Wigner parameterization (similar to (\ref{eq:uim-3})).

An extensive database of pion photoproduction experiments has been 
analyzed by SAID.  The extension of SAID to the analysis of pion 
electroproduction data is being pursued. 

\subsubsection{Giessen and KVI Models}
The K-matrix  coupled-channel models developed by 
the Giessen group~\cite{giessen1,giessen2,giessen3,giessen4} 
and the KVI group~\cite{kvi}
can be obtained from (\ref{eq:t-k}) by taking the approximation
$K = V$. This leads to a matrix equation involving only the
on-shell matrix elements of $V$
\begin{eqnarray}
T_{a,b}(E) \rightarrow
\sum_{c}[(1+i V(E))^{-1}]_{a,c} V_{c,b}(E) \,.
\label{eq:giessen}
\end{eqnarray}
The interaction $V=v^{bg} +v^R$ is calculated from tree diagrams  
such as those illustrated in Figure~\ref{fig:mechanism_1}.
The Giessen group has performed 
analyses~\cite{giessen1,giessen2,giessen3,giessen4} with
$\gamma N$, $\pi N$, $2\pi N$, $\eta N$, $\omega N$, and $K\Lambda$ channels.
They further simplify $2\pi N$ as a stable partice channel.  
The KVI group~\cite{kvi} has focused on the hyperon production reactions by
performing analyses with $\gamma N, \pi N, \eta N, K\Lambda, K\Sigma, \phi N$
channels.

\subsection{Dynamical Models}

The dynamical models of meson-baryon reactions account for the 
off-shell scattering dynamics through the use of
integral equations such as (\ref{eq:tmatrix}) or their equivalence
(\ref{eq:t-2pot})-(\ref{eq:nstar-sigma}). The off-shell dynamics
are closely related to the meson-baryon scattering wavefunctions in the
short-range region where we want to map out the structure of $N^*$.
Thus the development of dynamical models is an important step
towards interpreting the extracted $N^*$ parameters.

In recent years, the predictions from the dynamical models
of Sato and Lee (SL)~\cite{sl} and the
 Dubna-Mainz-Taiwan (DMT) collaboration~\cite{dmt} are most often used
to analyze the data in the $\Delta$ region.
The SL model can be derived from
(\ref{eq:t-2pot})-(\ref{eq:nstar-sigma}) by
keeping only one resonance $N^*=\Delta$ and 
two channels $a,b= \pi N, \gamma N$.
In solving exactly the resulting equations
the non-resonant interactions $v^{bg}_{\pi N,\pi N}$ and
$v^{bg}_{\pi N,\gamma N}$  are derived from the standard PV
Born terms 
and $\rho$ and $\omega$ exchanges by using an unitary transformation method. 

The DMT model also includes only 
$\pi N$ and $\gamma N$ channels. They however 
 depart from the exact formulation based on (\ref{eq:tmatrix}) or
(\ref{eq:t-2pot})-(\ref{eq:nstar-sigma})
by using the Walker parameterization (\ref{eq:uim-3}) to describe the resonant 
amplitude. Accordingly,
their definition of the non-resonant amplitude also differs from 
that defined by (\ref{eq:t-nonr}) : 
$t^{bg}_{c,b}$ in the right-hand side of (\ref{eq:t-nonr}) 
is replaced by the
 full amplitude $T_{c,b}$.
Furthermore, they follow MAID to calculate the non-resonant interaction
$v^{bg}_{\pi N,\gamma N}$ from an energy-dependent mixture of PS and PV
Born terms. 

Extensive data of pion photoproduction and electroproduction in 
the $\Delta$ region can be described by both the SL and DMT models. 
However, they have significant differences in the extracted 
electric E2 and Coulomb C2 form factors of the
 $\gamma N \rightarrow \Delta$ transition.
Both models show very large  pion cloud effects on the
 $\gamma N \rightarrow \Delta$ transition form factors in the  
low $Q^2$ region. 

The Utrecht-Ohio model~\cite{ohio} has also succeeded in
describing the data in the $\Delta$ region. Despite some differences in
treating the gauge invariance problem and the $\Delta$ excitation amplitude,
its dynamical content is similar to that of SL and DMT models.

The equations (\ref{eq:t-2pot})-(\ref{eq:nstar-sigma})
are used in a 2-$N^*$ and 3-channels ($\pi N$, $\eta N$,
and $\pi \Delta$)  
study~\cite{yosh} of $\pi N$ scattering in $S_{11}$ partial wave.
This work illustrated the extent to which
the  quark-quark interactions in
the constituent quark model can be determined directly
by the $\pi N$ reaction data.
The equations (\ref{eq:t-2pot})-(\ref{eq:nstar-sigma})
have also been used to show that
the  coupled-channel effects due to the
$\pi N$ channel are very large in  
$\omega$  photoproduction~\cite{ohlee} and 
$K$ photoproduction~\cite{chitab}. All $N^*$ identified by the
Particle Data Group are included in these two dynamical 
calculations.

The coupled-channel study of both $\pi N$ scattering and $\gamma N \rightarrow \pi N$
in the $S_{11}$ channel by Chen et al~\cite{chen} includes
$\pi N$, $\eta N$, and $\gamma N$ channels.
Their $\pi N$ scattering calculation is performed by using (\ref{eq:tmatrix}).
For their $\gamma N \rightarrow \pi N$ calculation, they neglect 
the $\gamma N \rightarrow \eta N \rightarrow \pi N $ channel coupling, 
and follow the procedure of the DMT model to 
 evaluate the resonant term in terms of
the Walker parameterization  (\ref{eq:uim-3}).
  They find that four
$N^*$  are needed to fit the
empirical amplitudes in the $S_{11}$ channel up to $W = 2$ GeV. 
This approach is being extended
 to also fit the $\pi N\rightarrow \pi N$ and
$\gamma N\rightarrow \pi N$ amplitudes in all
partial waves. 

A coupled-channel calculation based on (\ref{eq:tmatrix}) 
has been carried out
by the J\"ulich group~\cite{julich} for $\pi N$ scattering.
They are able to describe the $\pi N$ phase shifts
up to $W=1.9$ GeV by including the $\pi N$, $\eta N$,
$\pi \Delta$, $\rho N$ and $\sigma N$ channels and 5 $N^*$ resonances:~$P_{33}(1232)$, $S_{11}(1535)$, 
$S_{11}(1530)$, $S_{13}(1650)$ and $D_{13}(1520)$. They find that the Roper resonance
$P_{11}(1440)$ arises completely from the meson-exchange coupled-channel effects.

A coupled
channel calculation based on (\ref{eq:tmatrix}) for both the
$\pi N$ scattering and $\gamma N \rightarrow \pi N$ up to $W=1.5$ GeV has
been reported by Fuda and Alarbi~\cite{fuda}. 
 They include the $\pi N$, $\gamma N$, $\eta N$ and $\pi \Delta$
channels and 4 $N^*$ resonances:~$P_{33}(1232)$, $P_{11}(1440)$, $S_{11}(1535)$
and $D_{13}(1520)$. They adjust the parameters of their model
to fit the empirical multipole amplitudes in
low partial waves.

More simple coupled-channel calculations have been  performed by using 
separable interactions. In the model of Gross and Surya~\cite{gros}, separable 
interactions come from simplifying the meson-exchange mechanisms 
in Figs~\ref{fig:mechanism_1}(a)-\ref{fig:mechanism_1}(c) 
as a contact term
like Figure~\ref{fig:mechanism_1}(d). They include only $\pi N$ and 
$\gamma N$ channels  and 3 resonances:~$P_{33}(1232)$, $P_{11}(1440)$ and $D_{13}(1520)$,
and restrict their investigation up to 
$W < 1.5$ GeV. To account for the inelasticities in $P_{11}$ and $D_{13}$, the $N^*\rightarrow
\pi\Delta$ coupling is introduced in these two partial waves. The inelasticities in other
partial waves are neglected.

A similar separable simplification is also used in 
the chiral coupled-channel models~\cite{kais1,oset1} for strange 
particle production. There the separable interactions
 are directly deduced from
the SU(3) effective chiral lagrangians.
They are able to fit the total cross section data for various strange
particle production reaction channels without introducing $N^*$
resonance states.  In recent years, this model has been further extended by Lutz
and  Kolomeitsev~\cite{lutz} to also fit the $\pi N$ scattering data.

None of the dynamical models described above account for all of the effects arising from
the $\pi\pi N$ channels, which contribute about 1/2 of the $\pi N$ and $\gamma N$ total
cross sections in the higher mass $N^*$ region.
Consequently, the $N^*$ parameters extracted from these
models could have uncertainties due to the violation of the
$\pi\pi N$ unitarity condition.
One straightforward way to improve the situation
is to extend the Hamiltonian formulation
of \cite{sl} to include (a) 
$\rho \rightarrow \pi\pi$  and $\sigma \rightarrow \pi\pi$ decay
mechanisms as specified in
$\Gamma_V$ of (\ref{eq:fv}), (b)
$v_{\pi\pi}$ for non-resonant $\pi\pi$ interactions, (c)
$v_{MN, \pi\pi N}$ for non-resonant
$M N \rightarrow \pi\pi N$ transitions with $M N =\gamma N$ or $\pi N$,
and (d) $v_{\pi\pi N,\pi\pi N}$ for non-resonant 
$\pi\pi N \rightarrow \pi\pi N$ 
interactions. Such a dynamical
coupled-channel model has been developed recently by 
Matsuyama, Sato, and Lee~\cite{msl} (MSL).
The coupled-channel equations from this model can also be cast into
the form of (\ref{eq:t-2pot})-(\ref{eq:nstar-sigma})
except that the driving term of (\ref{eq:t-nonr}) is replaced by
\begin{eqnarray}
v^{bg}_{a,b} \rightarrow \hat{V}_{a,b} = v^{bg}_{a,b}
+Z_{a,b}(E) \,,
\label{eq:new-vbg}
\end{eqnarray}
where
\begin{eqnarray}
Z_{a,b}(E)
&=&   < a \mid F
\frac{ P_{\pi \pi N}}
{E- H_0 - v_{\pi N,\pi N}- v_{\pi\pi}
 - v_{\pi\pi N,\pi\pi N} + i\epsilon} F^\dagger \mid b >  \nonumber \\
& & -<a|\Sigma_V(E)|b>\delta_{a,b}
\label{eq:pipin-int}
\end{eqnarray}
with
\begin{eqnarray}
F&=&\Gamma_V + v_{\pi N, \pi\pi N} \nonumber \\
&=&\Gamma_{\Delta,\pi N}+h_{\rho,\pi\pi}+
h_{\sigma,\pi\pi}+v_{\pi N, \pi\pi N} \,.
\label{eq:fv}
\end{eqnarray}
 Note that $\Sigma_V(E)$ in (\ref{eq:pipin-int})
describes the self-energy
of unstable particle channels
$\pi \Delta$, $\rho N$, and $\sigma N$.
Obviously $Z_{a,b}(E)$ contains the non-resonant multiple scattering
mechanisms within the $\pi\pi N$ subspace.
It generates  $\pi\pi N$ unitarity cuts.
It has been shown in \cite{msl} that this term has very large effects
on $2\pi$ production reactions. This dynamical coupled-channel model is
currently being further developed at the Excited Baryon Analysis Center (EBAC)
at JLab.

\section{Experimental developments}

The past ten years has seen an unprecedented increase in the volume and 
precision of $N^*$ data from electromagnetic meson production reactions.  This 
has resulted from the development of CW electron accelerators, which 
drastically reduce accidental backgrounds in coincidence experiments, as 
well as from advances in charged, neutron and photon detector technologies 
such as nearly $4\pi$ solid angle coverage and improved spatial/energy 
resolution.  In addition, the development of polarized beams, targets 
and polarimeters has opened up a previously unavailable set of observables 
which are sensitive to the interference between resonant and non-resonant
processes in the scattering amplitude.  An extensive survey of initial
data from the new facilities can be found in the review by Burkert and 
Lee~\cite{lee-review}.  In this section we highlight results from 
recent analyses of ongoing experiments.

\subsection{Resonance transition form factors from $\pi$, $\eta$, $2\pi$ production data}
Pion electroproduction provides new information not available using real photons.  
First, measuring the $Q^2$ dependence of the photocoupling amplitudes probes 
a range of distance scales within the nucleon, revealing spatial information 
about the constituents. Second,  polarization of the virtual 
photon allows study of both longitudinal and transverse modes of resonance 
excitation.  Both of these factors place strong constraints on the structure
of the transition current used in phenomenological, quark model and 
lattice QCD calculations.
\subsubsection{$\Delta(1232)$ Resonance}

A longstanding question is the origin of the non-zero quadrupole strength
experimentally observed in the $\gamma^* p\rightarrow \Delta(1232)$ 
transition.  Quadrupole transitions, which occur via the absorption of $E2$ or 
$C2$ photons, imply non-zero orbital angular momentum in the $N$ or $\Delta$ 
wave function, which can arise from configuration mixing due to residual 
gluon or meson interactions between constituent quarks or from the direct photon 
coupling to the pion cloud. 
Until recently the data were not sufficiently accurate to constrain calculations.
\begin{figure} 
\centerline{\includegraphics[width=12cm]{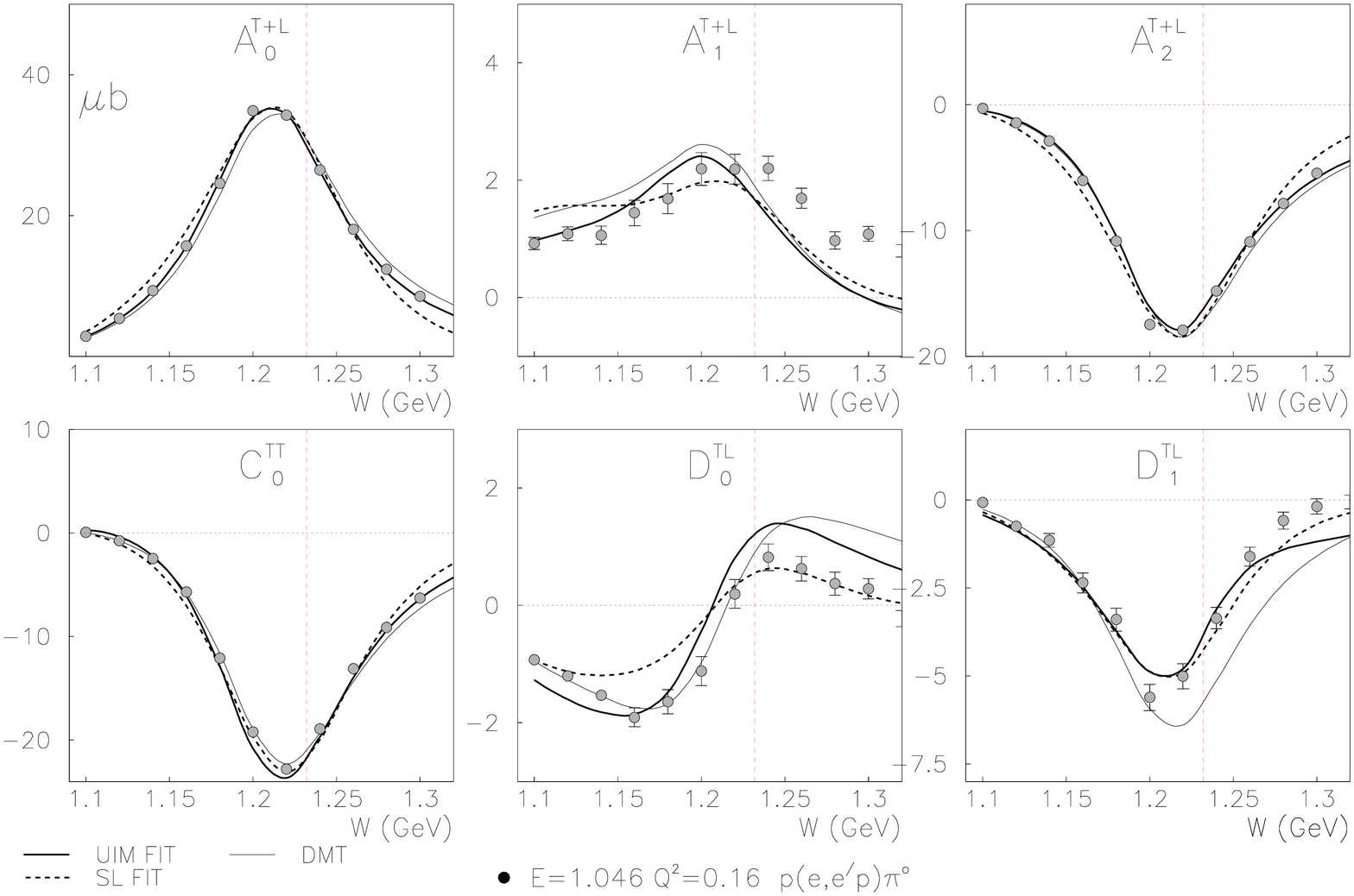}}
\caption{$W$ dependence of partial wave coefficients from Legendre fit of 
experimental $\pi^0$ electroproduction structure functions 
$\sigma_{T+L}$, $\sigma_{TT}$ and $\sigma_{TL}$.  Solid curve shows UIM 
fit results.  Vertical line shows position of $\Delta(1232)$ resonance.  
CLAS data~\protect\cite{Smi06} taken at E=1.046 GeV and $Q^2=0.16$~GeV$^2$.}
\label{fig:uimfit}
\end{figure}
New precision measurements of the $p(e,e^{\prime}p)\pi^o$ reaction at 
BATES, JLAB and MAMI have drastically improved our knowledge of the 
$Q^2$ dependence of the electric and Coulomb quadrupole/magnetic dipole 
ratios $R_{EM}=E_{1+}/M_{1+}$ and $R_{SM}=S_{1+}/M_{1+}$.  The large acceptance 
CLAS experiments at JLAB measure the interference terms $Re(E_{1+}M^*_{1+})$ 
and $Re(S_{1+}M^*_{1+})$ via the azimuthal and partial wave decomposition 
of the $p\pi^0$ final state:
\begin{eqnarray}
\frac{d\sigma}{d\Omega^*_{\pi}} = \sum_{\ell=0}^{2}A_\ell^{T+L} P_\ell(\cos\theta^*_{\pi})
\,&+&\,\sin^2\theta^*_{\pi}\,C_{0}^{TT}\cos 2\phi^*_{\pi}  \nonumber\\
\,&+&\,\sin\theta^*_{\pi}\sum_{\ell=0}^{1} D_\ell^{TL} P_\ell(\cos\theta^*_{\pi})\cos\phi^*_{\pi} 
\label{eq:1}
\end{eqnarray}
where to first order, retaining only $M_{1+}$ dominated terms
\begin{eqnarray} 
|M_{1+}|^2 &=& A_{\,0}^{\,T+L}/2,  \label{eq:29} \\
Re(E_{1+}M^*_{1+}) &=& (A_{\,2}^{T+L} -2\,C_{\,0}^{\,TT}/\,3)/8, \label{eq:30} \\
Re(S_{1+}M^*_{1+}) &=& D_{\,1}^{\,TL}/6.\label{eq:31}
\end{eqnarray}
Here "T" and "L" refer to the transverse and longitudinal components of the
virtual photon polarization.  The data are shown in Figure~\ref{fig:uimfit} 
for $Q^2=0.2$~GeV$^2$ where it is 
seen that all the Legendre coefficients in (\ref{eq:1}) show a $W$ dependence 
consistent with interference terms involving the strong resonant $M_{1+}$.  
At higher $Q^2$, where $M_{1+}$ dominance is no longer valid, corrections 
to (\ref{eq:29}-\ref{eq:31}) can be made using the Unitary Isobar Model (UIM) 
described in Section 2.1 where non-resonant backgrounds and tails from 
higher resonances are parameterized using (\ref{eq:uim-2}) and (\ref{eq:uim-3}). 
A typical UIM fit is shown in Figure~\ref{fig:uimfit}.

\begin{figure} 
\centerline{\includegraphics[width=10cm]{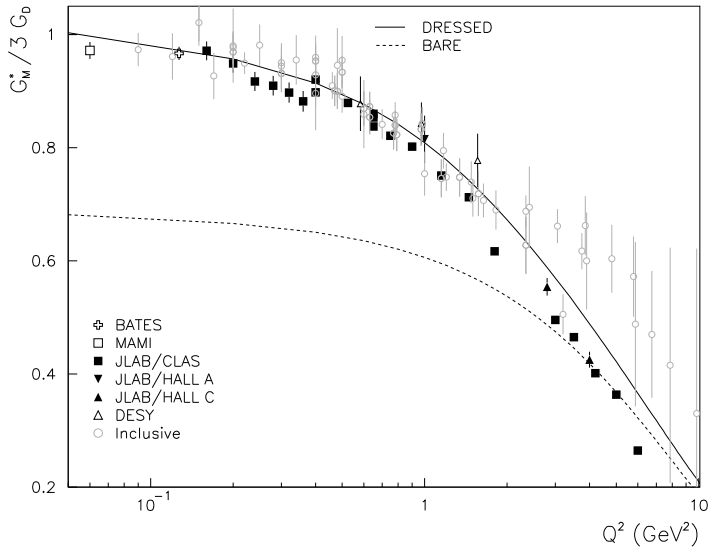}}
\caption{Magnetic dipole transition form factor $G^*_M$ 
for $\gamma^* N \rightarrow \Delta(1232)$, normalized to 
the elastic dipole $G_d$.  Inclusive data are from pre-1990 
experiments at Bonn, DESY and SLAC. Other data points were 
obtained from Unitary Isobar Model analysis of the exclusive 
$p(e,e^{\prime}p)\pi^o$ reaction.  Curves from Sato-Lee 
model discussed in text.}
\label{fig:gm_Delta}
\end{figure}

\begin{figure}
\centerline{\includegraphics[width=8cm]{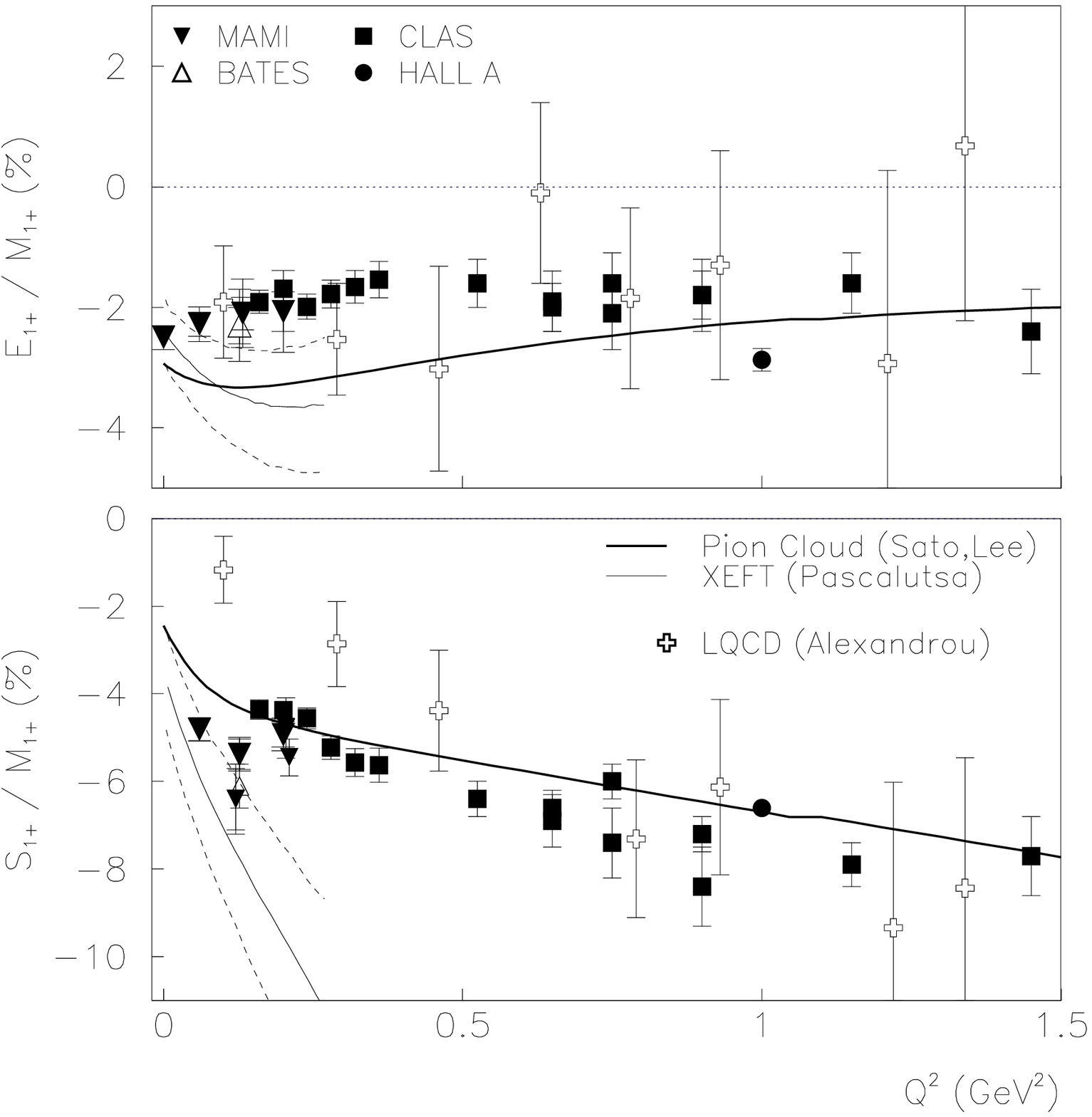}
            \includegraphics[width=4.36cm]{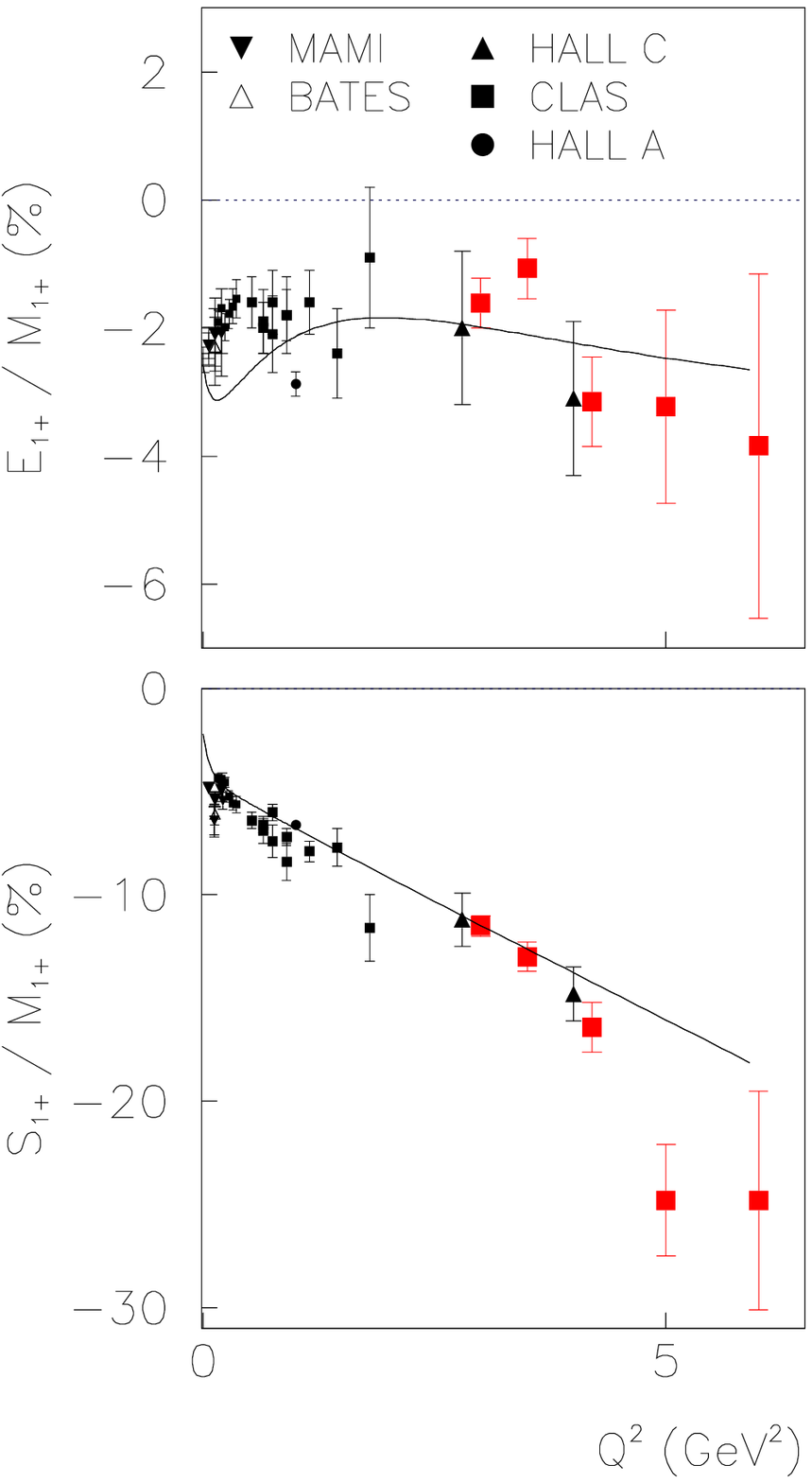}}
\caption{Quadrupole $\gamma N \rightarrow \Delta(1232)$ transition 
form factors at low $Q^2$ (left) and high $Q^2$ (right) from recent
experiments~\protect\cite{lee-review}. Electric quadrupole $E_{1+}$ 
and scalar quadrupole $S_{1+}$ are normalized to the magnetic dipole $M_{1+}$.  
Preliminary data from CLAS~\protect\cite{Smi06,Ung06} and MAMI~\protect\cite{Sta06,Spa06} 
are also included.}
\label{fig:mrat_delta}
\end{figure}

Figure~\ref{fig:gm_Delta} shows the $Q^2$ dependence of the magnetic dipole $G_M^*$, 
which is related to the isospin $I=3/2$ $\pi N$ multipole $Im(M_{1+}^{3/2})$ 
through the $\Delta$ width and a kinematic factor.  These new exclusive measurements
of $G_M^*$ now extend over the range $Q^2=0.06-6.0$~GeV$^2$, and confirm the rapid 
$Q^2$ falloff relative to the elastic dipole seen in previous inclusive measurements,
but with much greater sensitivity to the resonant $M_{1}+$ multipole at higher $Q^2$.
As previously noted, at least part of this rapid falloff may arise from 
the 'dressing' of the $\gamma^*N\rightarrow \Delta$ vertex by rescattering through 
the pion cloud.  The 'bare' curve in Figure~\ref{fig:gm_Delta} is obtained
in the SL dynamical model when the predicted 'dressed' $G_M^*$ form factor is 
fitted to the data.  It shows the increasing contribution from mesons 
as $Q^2\rightarrow 0$.  Similar results have been obtained in the 
Dubna-Mainz-Taipei model.

\begin{figure}
\centerline{\includegraphics[width=7cm,angle=90]{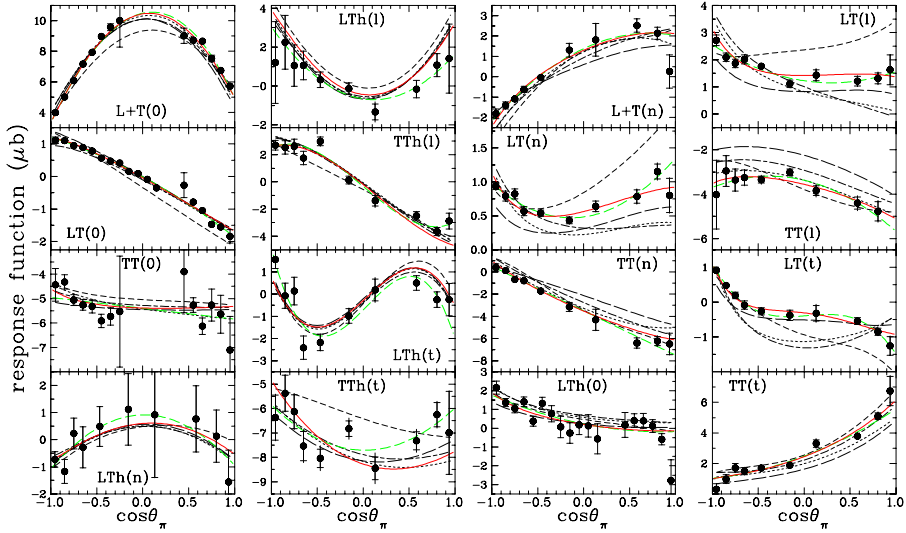}}
\caption{Response functions for $p(\vec{e},e\,\vec{p}^{\prime})\pi^0$ measured
by Kelly $\it{et al}$~\protect\cite{kel05a,kel05b} at $W=1.23$~GeV and $Q^2=1.0$~GeV$^2$.  
Notation refers to transverse~(t), normal~(n) or longitudinal~(l) components 
of the proton recoil polarization.  MAID, DMT, SAID and SL models given by 
dash-dotted, dotted, short-dashed and long-dashed curves.}
\label{fig:kelly}
\end{figure}

The experimental quadrupole transition ratios REM and RSM are shown in 
Figure~\ref{fig:mrat_delta}.  The JLAB results are based on a full 
partial-wave analysis, while the MAMI and BATES data sets cover a
more limited angle range.  For the latter, MAID, DMT and SL model fits are 
used to estimate the $\Delta$ resonant multipoles.  The other data points 
with errors are quenched lattice QCD calculations~\cite{dina} which have undergone a 
linear chiral extrapolation to the physical pion mass.  It is evident that
while the calculations qualitatively account for the magnitude of
REM and RSM, the $Q^2$ dependence is not consistently described.  
The low $Q^2$ underprediction of RSM by LQCD is particularly striking.  It
was recently demonstrated, using a relativistic chiral effective field theory 
($\chi$EFT) calculation~\cite{pasc-3}, that RSM may be particularly 
sensitive to a non-analytic dependence of pion loop diagrams on the quark mass, 
rendering invalid the linear chiral extrapolation used in~\cite{dina}.  A strong negative
slope near $Q^2=0$ is predicted by $\chi$EFT for both REM and RSM 
(see Figure~\ref{fig:mrat_delta}).  While the lowest $Q^2$ RSM points from 
MAMI and BATES appear to agree with this prediction, the data for higher
$Q^2$ are in better agreement with the SL and LQCD predictions.  For REM,
the data do not show the strong $Q^2$ dependence predicted by $\chi$EFT
near $Q^2=0$, and also lie outside of the estimated theoretical 
uncertainty arising from higher-order effects. 

At the largest $Q^2$ measured by CLAS (right panel in Figure~\ref{fig:mrat_delta}) 
neither REM nor RSM exhibit the helicity conserving scaling behavior 
expected in the hard pQCD limit (REM$\rightarrow 1$, RSM$\rightarrow constant$).  
Normally the transverse momenta of quarks are expected to be of
the order of $\Lambda_{QCD}$, and thus only important for resonance excitations
at low $Q^2$.  However Idilbi, Ji and Ma~\cite{Ji04} have shown that the orbital 
motion of small$-x$ partons makes
a non-neglible contribution to helicity non-conservation in the $N\Delta$
transition, thus invoking both soft and hard QCD processes. 

\begin{figure} 
\centerline{\includegraphics[width=8cm]{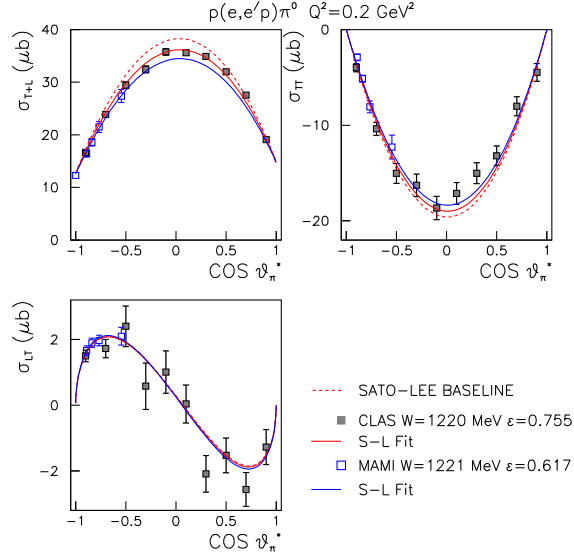}}
\caption{Fits of Sato-Lee dynamical model to experimental $p(e,e^{\prime}p)\pi^0$
structure functions $\sigma_{T+L}$, $\sigma_{TT}$ and $\sigma_{LT}$.  Preliminary
data from CLAS (solid)\protect~\cite{Smi06} and MAMI (open)\protect~\cite{Spa06} 
are shown at $Q^2=0.2$~GeV$^2$.  The 'baseline' curve refers to the default
$Q^2$ parameterization of the 'bare' couplings $g_m, g_e$ and $g_c$ prior to the fit.}
\label{fig:slfit1}
\end{figure}

\begin{figure}
\centerline{\includegraphics[width=8cm]{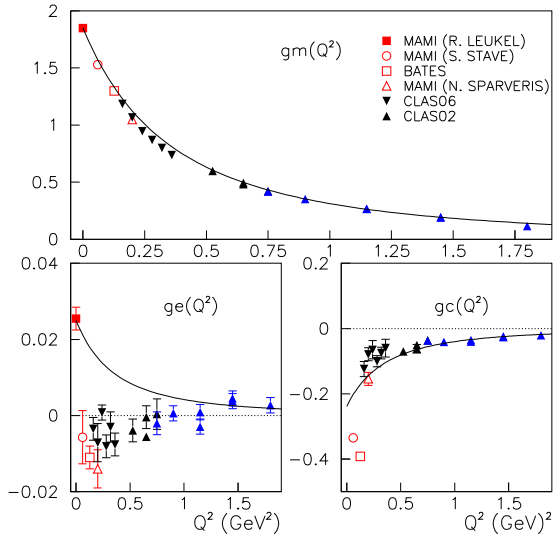}}
\caption{$Q^2$ dependence of bare magnetic dipole, electric quadrupole and coulomb
quadrupole couplings $g_m, g_e$ and $g_c$ for the $\gamma^*p\rightarrow \Delta$ transition,
extracted from fits of SL dynamical model to BATES, CLAS and MAMI data.  
Solid line assumes $Q^2$ dependence of $g_e$ and $g_c$ couplings are identical to $g_m$.}
\label{fig:slfit2}
\end{figure}

In addition to understanding the long-range chiral dynamics which are evidently
important near $Q^2=0$, there has long been a fundamental interest in the 'bare' 
$\gamma N\rightarrow \Delta$ vertex, which is presumably more sensitive to 
the short-range interquark potential relevant to QCD-motivated models.  
The bare couplings, which also contain information about the quark core 
wave functions of the $N$ and $\Delta$, can be estimated
in dynamical reaction models by treating them as free parameters to be
fitted to the experimental data.  This was performed using the SL model for
all recent data sets below $Q^2=2.0$~GeV$^2$.  A typical fit to CLAS and MAMI
data at $Q^2=0.2$~GeV$^2$ is shown in Figure~\ref{fig:slfit1}, while the 
$Q^2$ dependence of the extracted bare couplings $g_m, g_e$ and $g_c$ are 
shown in Figure~\ref{fig:slfit2}.  The simplest assumption, that the bare electromagnetic
couplings have an identical $Q^2$ dependence, is clearly ruled out by the
data.  Unlike $g_m$, both $g_e$ and $g_c$ show a strong departure from smooth
behavior at low $Q^2$.  This behavior may indicate the presence of additional
long-range physics not included in the reaction models.  It is certain that 
more experimental data at the lowest $Q^2$ would aid the interpretation of
these results.

All reaction models make implicit assumptions about the mechanisms responsible
for non-resonant backgrounds.  This often leads to substantial model dependence
in the extraction of resonance parameters due to ambiguities in relative phases,
omitted diagrams, etc.  In addition these same models are often used to perform the
partial wave analysis of the experimental data, which should in principle be a model
independent procedure.  Clearly, more complete experiments would help resolve these
ambiguities.  A recent JLAB/Hall A experiment used a polarized beam and a high
resolution magnetic spectrometer instrumented with a recoil polarimeter to measure 
16 neutral pion electroproduction response functions at $Q^2$=1.0~GeV$^2$.
Twelve of these response functions were observed for the first time. These data are
shown in Figure~\ref{fig:kelly} at the peak of the $\Delta(1232)$, compared to 
MAID, DMT, SAID and SL model predictions.  Although the overall agreement is good,
there are several response functions for which the models clearly diverge in their
predictions.  This could be due simply to lack of previous experimental information
on the weaker underlying multipoles.      Due
to the phase sensitivity of these polarization observables, a nearly model-independent
extraction of $REM$ and $RSM$ was obtained.
 
\subsubsection{$P_{11}(1440)$}
\begin{figure}
\centerline{
  \begin{minipage}[c]{0.45\linewidth}
    \includegraphics[width=\columnwidth]{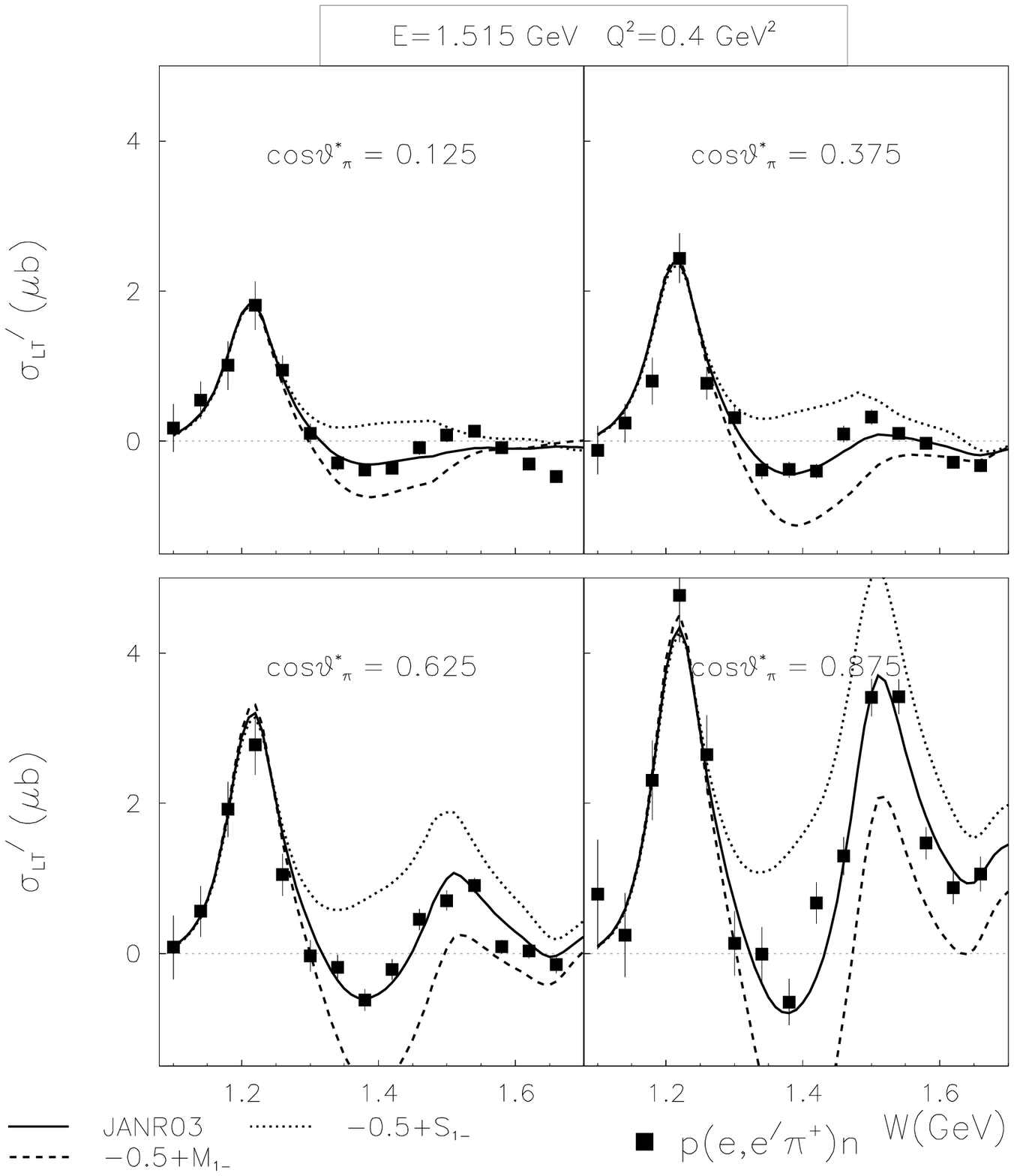}
  \end{minipage}\hfill
  \begin{minipage}[c]{0.5\linewidth}
    \includegraphics[width=\columnwidth]{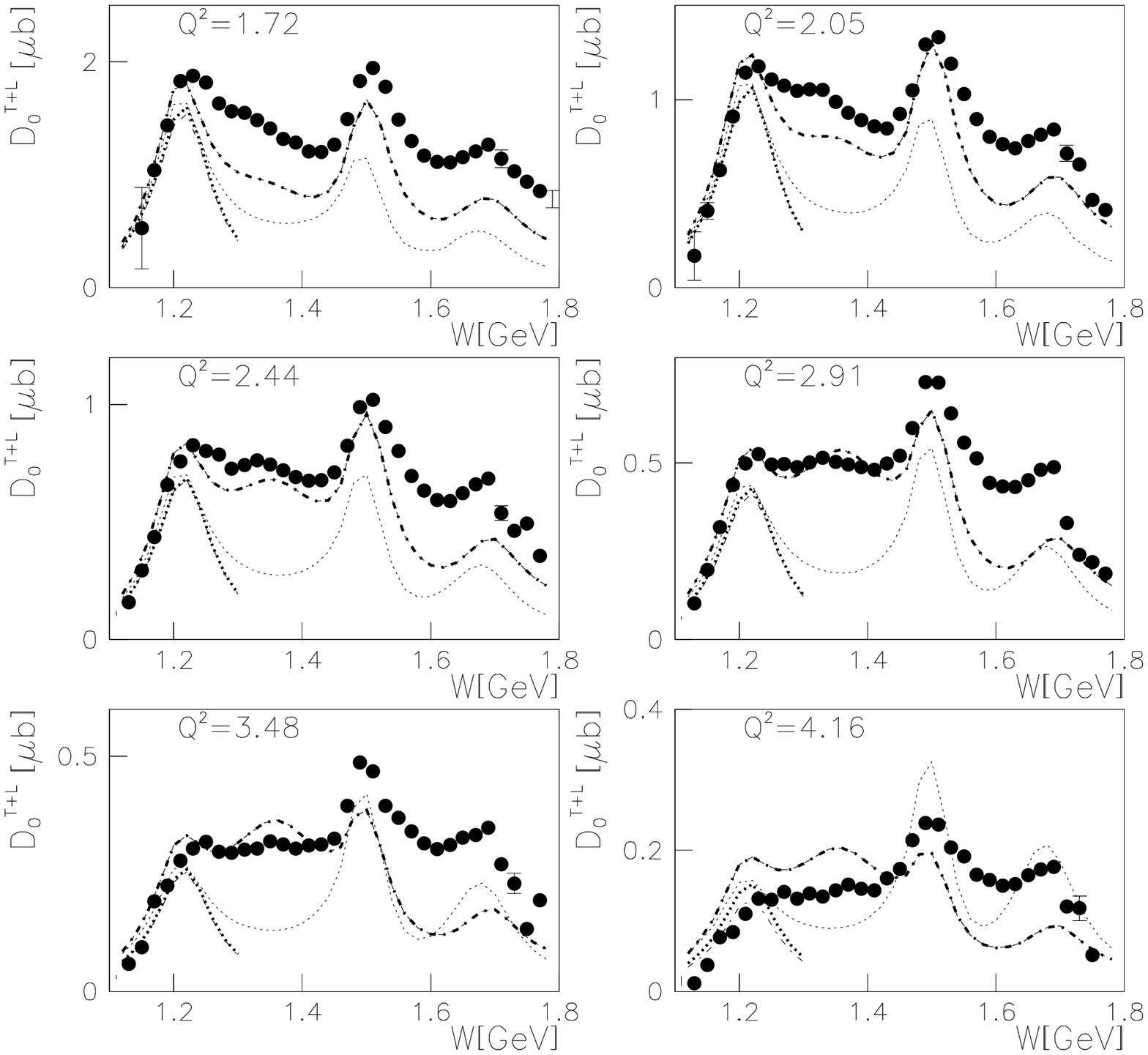}
  \end{minipage}
}
\caption{CLAS measurement of the $W$ dependence of $\sigma_{LT^{\prime}}$ (left)
and $\sigma_{T+L}$ (right) for the $p(e,e^{\prime}\pi^+)n$ reaction.  The curves
show UIM (left) and MAID (right) model sensitivity to changes in multipoles 
$M^{1/2}_{1-}$ and $S_{1-}^{1/2}$ involved in formation of the 
Roper resonance. Data were taken at $Q^2=0.4$~GeV$^2$ (left) and $Q^2=1.72-4.16$~GeV$^2$
(right).}
\label{fig:p11_sensitivity}
\end{figure}

A significant new result using the large acceptance data from CLAS is the 
demonstration~\cite{Joo05} of a large sensitivity to the $M^{1/2}_{1-}$ 
and $S^{1/2}_{1-}$ multipoles.  These multipoles receive 
large contributions from excitation of the $I=1/2$ $P_{11}(1440)$ or "Roper" resonance.
Due to the isospin coupling, this sensitivity is maximized in the $n\pi^+$ channel.  It turns 
out that UIM fits to combined data sets which include the helicity dependent
asymmetry $A_{LT}^{\prime}(n\pi^+)$ and the cross section $d\sigma(n\pi^+)$ taken in 
the second resonance region, significantly constrain the real and 
imaginary parts of these multipoles through their interference, based on the UIM 
analysis, with largely Born dominated non-resonant multipoles.  The sensitivity of
the quantity $\sigma_{LT}^{\prime}\approx d\sigma\cdot A_{LT}^{\prime}$ to the 
imaginary (resonant) part of these multipoles is demonstrated in the left panel of Figure~\ref{fig:p11_sensitivity}. The plot shows the sensitivity of this observable 
at $Q^2=0.4$~GeV$^2$ to a 0.5~$\mu b^{\,1/2}$ shift in the transverse and 
longitudinal resonant photocouplings. The sensitivity is the result of the interference term that mixes real and imaginary amplitudes 
\begin{eqnarray}
\sigma_{LT^{\prime}} \sim Im(L)\cdot Re(T) - Im(T)\cdot Re(L)~, 
\end{eqnarray}
where L and T represent the longitudinal and transverse amplitudes, respectively.
At much higher $Q^2$, the tail from the
$\Delta(1232)$ is strongly diminished, and the Roper is directly visible as a 
small bump in the total $n\pi^+$ cross section near $W=1.4$~GeV (right panel).  

\begin{figure} 
\centerline{\includegraphics[width=6.0cm]{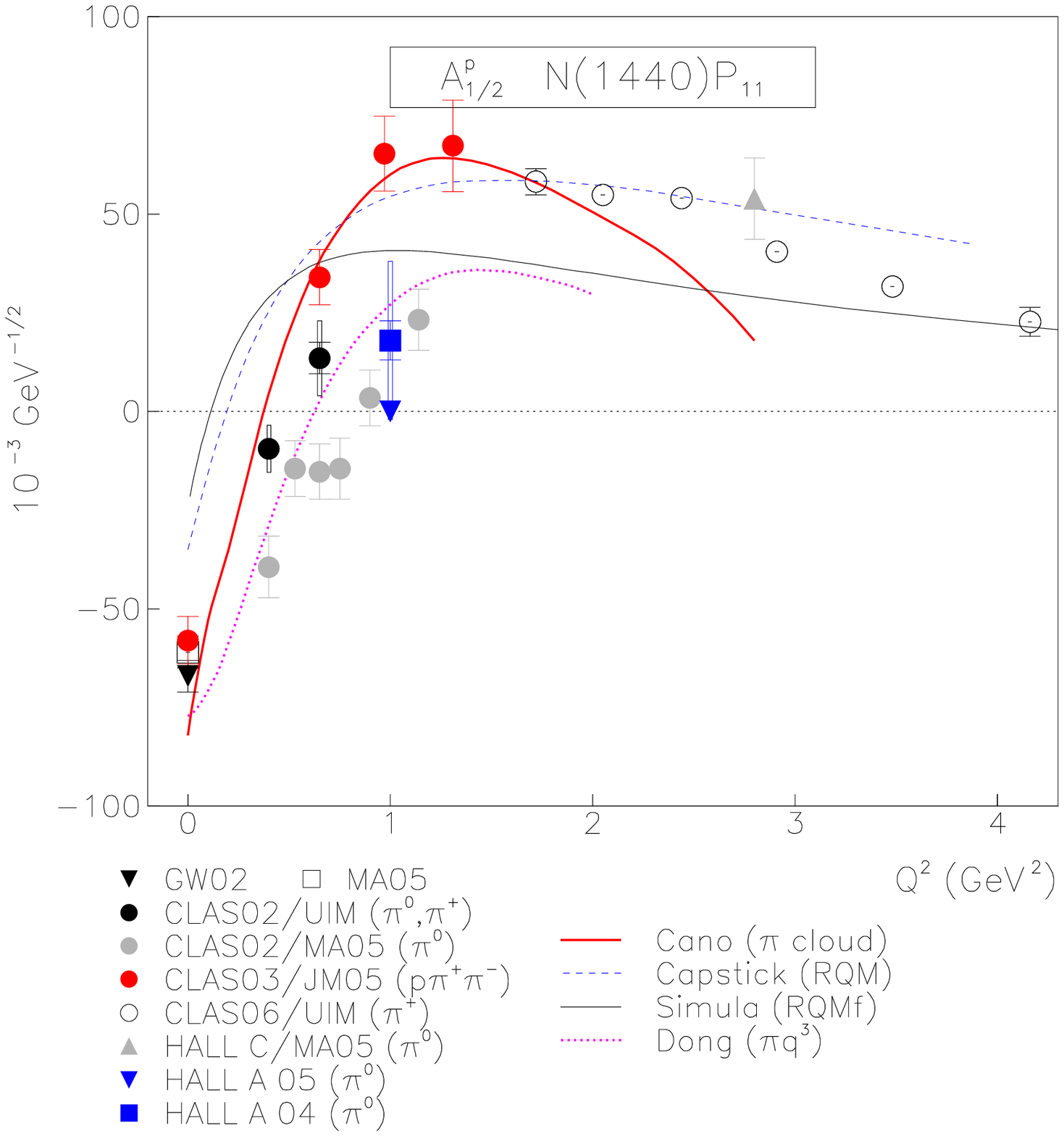}
	    \includegraphics[width=6.0cm]{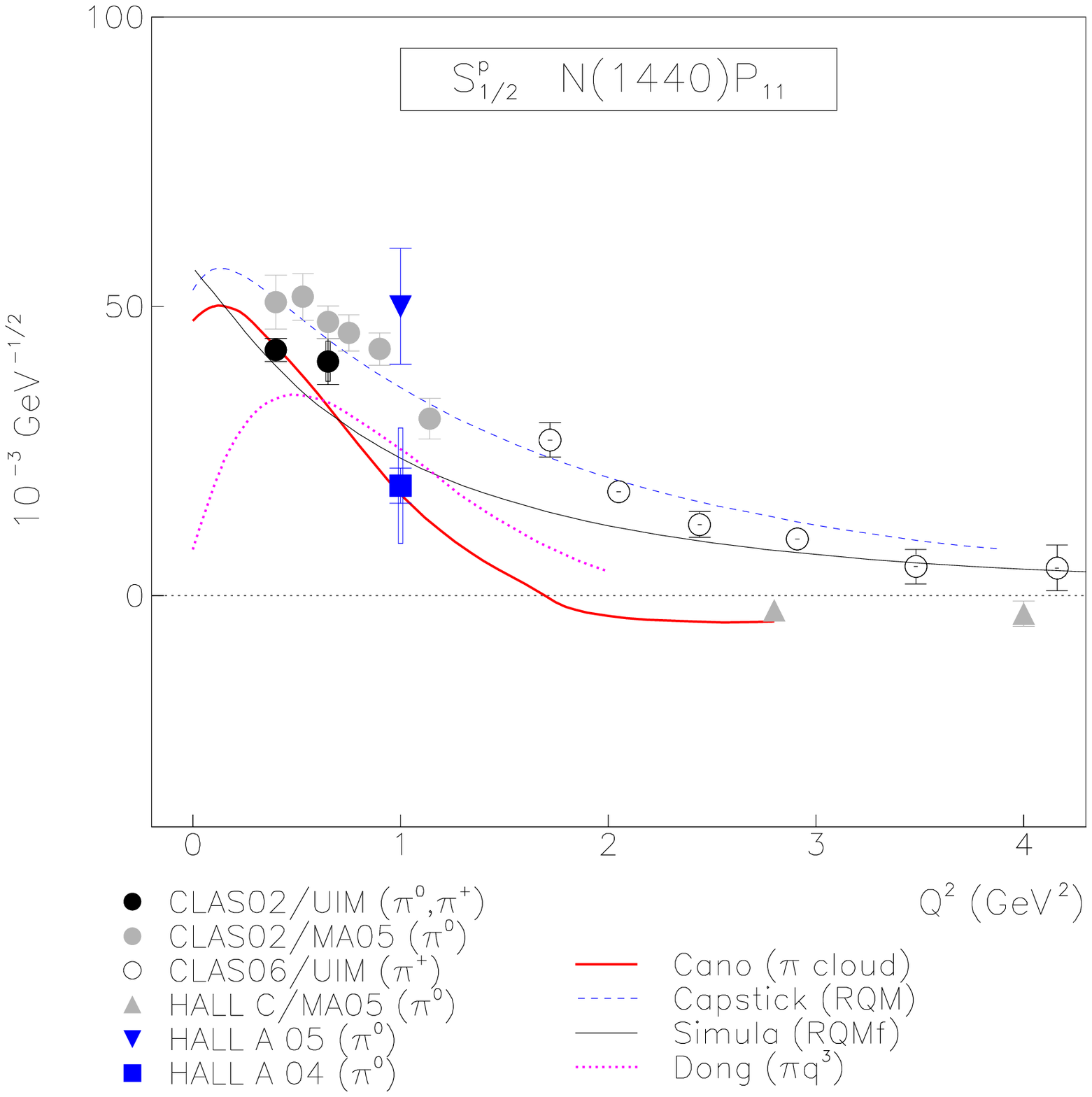}}
\caption{$Q^2$ evolution of the $P_{11}(1440)$ (Roper) amplitudes $A^p_{1/2}$ and $S^p_{1/2}$. 
Both CLAS/UIM and Hall A points include model error.  The MA05 analyses~\protect\cite{Tia04}
are dominated by $p\pi^0$ data.  Other points are GW02~\protect\cite{Arndt:1996}.  
The curves are model calculations \protect\cite{Can98,Cap95,Aie98,Pac99,War90,Don97} described in the text.}
\label{fig:p11}
\end{figure}

Using UIM and MAID model fits to CLAS data, the Roper photocoupling amplitudes 
have now been extracted over a large range of $Q^2$, as shown in Figure~\ref{fig:p11}.
A few of the $Q^2$ points have been further analyzed in a  multi-channel global fit~\cite{Azn05b}
which includes the $p\pi^0$ and $p\pi^+\pi^-$ final states and the agreement with
the single channel analysis is good. 
We also show recently published~\cite{Lav04,kel05b} points at $Q^2=1.0$~GeV$^2$ from Hall A. 
The new CLAS06 data at even higher $Q^2$~\cite{Par06} confirm the zero-crossing of $A^p_{1/2}$ 
first reported in~\cite{Azn05b}.  This feature can completely rule out some previous 
quark model calculations, such as most non-relativistic approaches, as well as the 
quark-gluon hybrid approach for which no zero crossing occurs and for which $S^p_{1/2}=0$.
The strong longitudinal response seen in these data could revive the traditional 
'breathing mode' description of the Roper, possibly augmented by coupling to the
$q\overline{q}$ cloud of the nucleon as suggested by the good agreement with the 
VMD model of Cano and Gonzales~\cite{Can98}.  Dynamical models and lattice QCD 
calculations could provide useful input towards intepreting the zero-crossing 
seen in $A^p_{1/2}$.  

\subsubsection{Second and Third Resonance Regions}
\begin{figure} 
\centerline{\includegraphics[width=6.0cm]{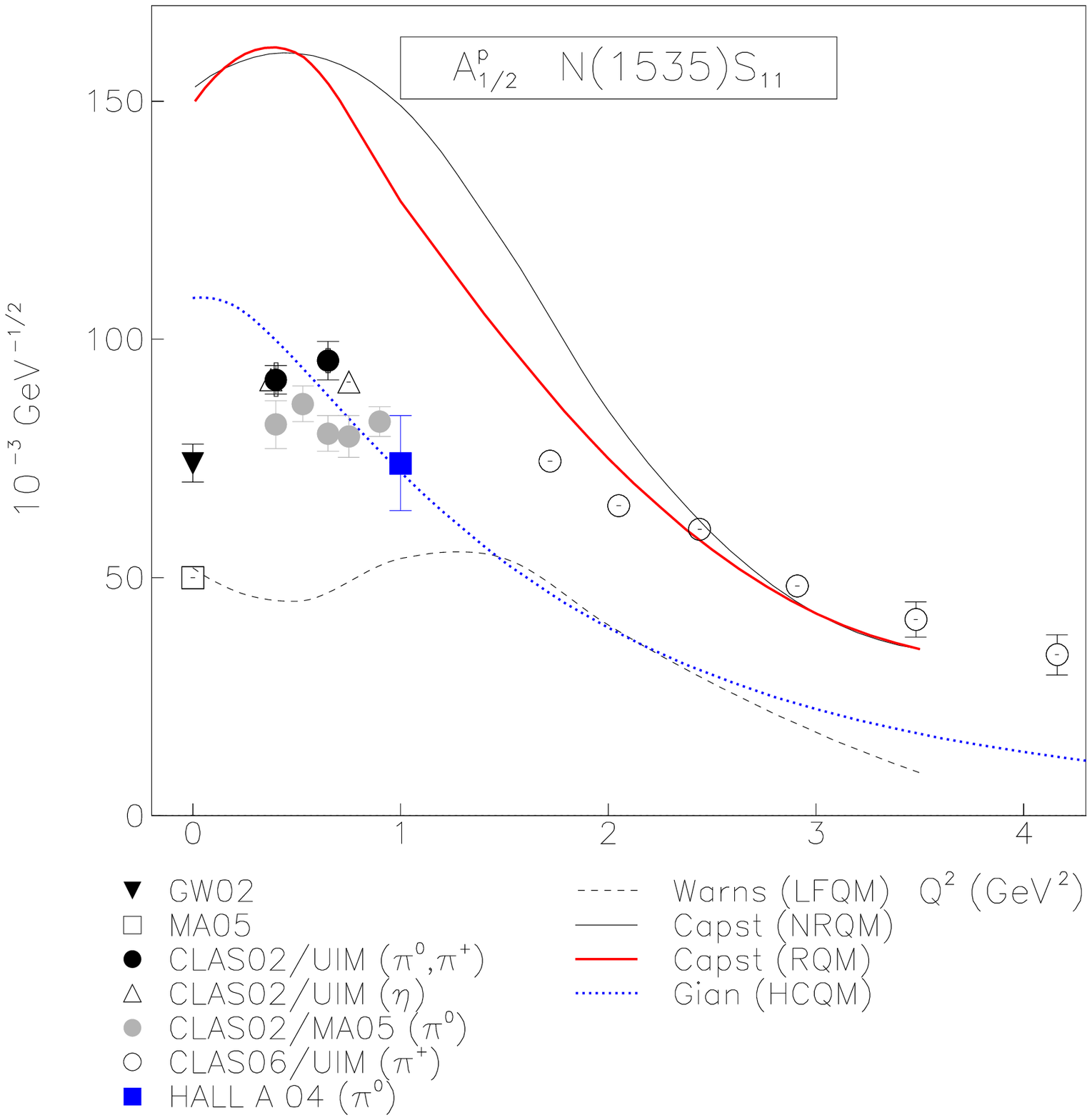}
            \includegraphics[width=6.0cm]{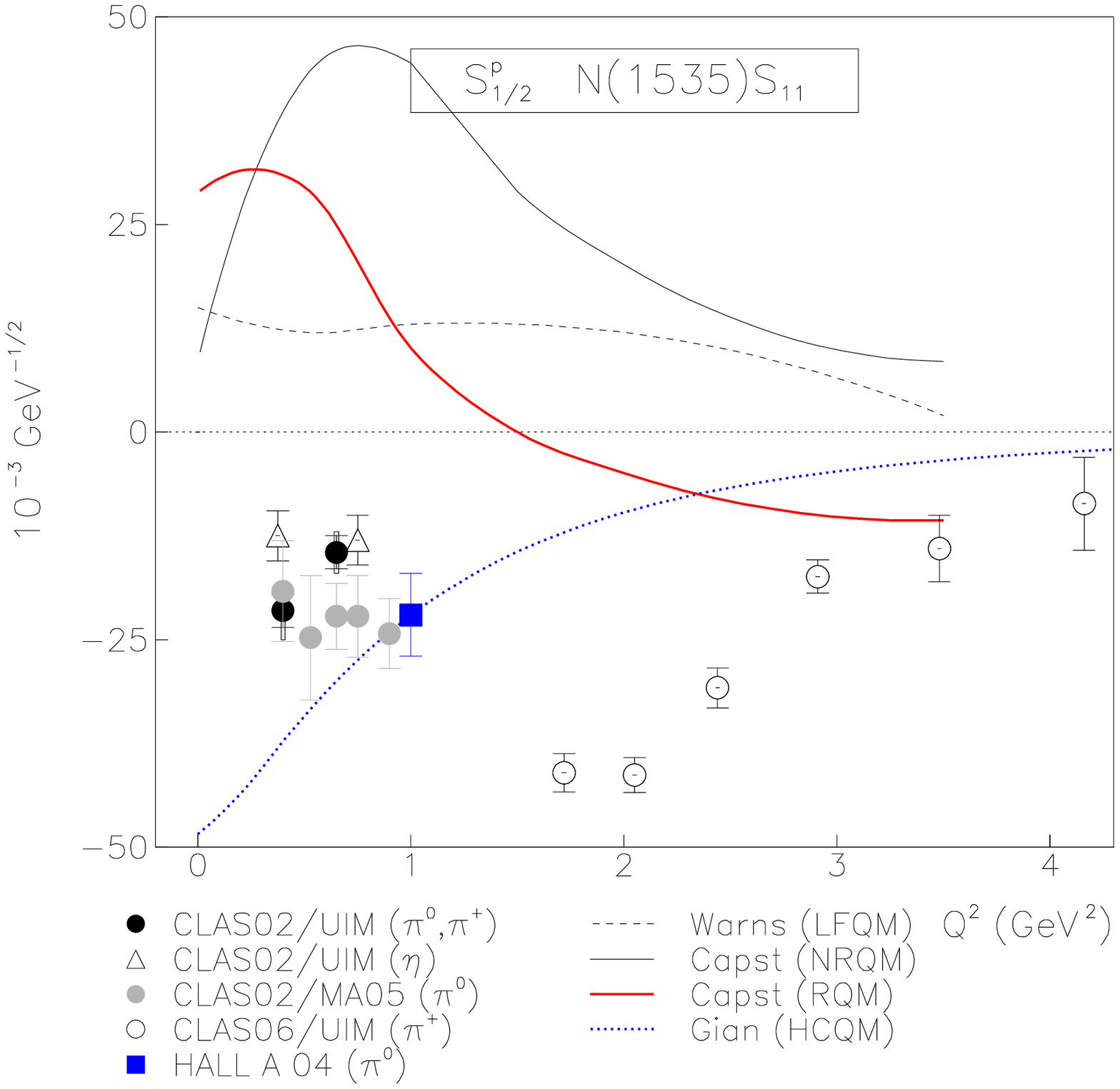}} 
\caption{$Q^2$ evolution of the $\gamma p \rightarrow S_{11}(1535)$ 
photocouplings amplitudes $A_{1/2}$ and $S_{1/2}$. See Figure~\ref{fig:p11}
for explantion of notation.}
\label{fig:s11}
\end{figure}

\begin{figure} 
\centerline{\includegraphics[width=6.0cm]{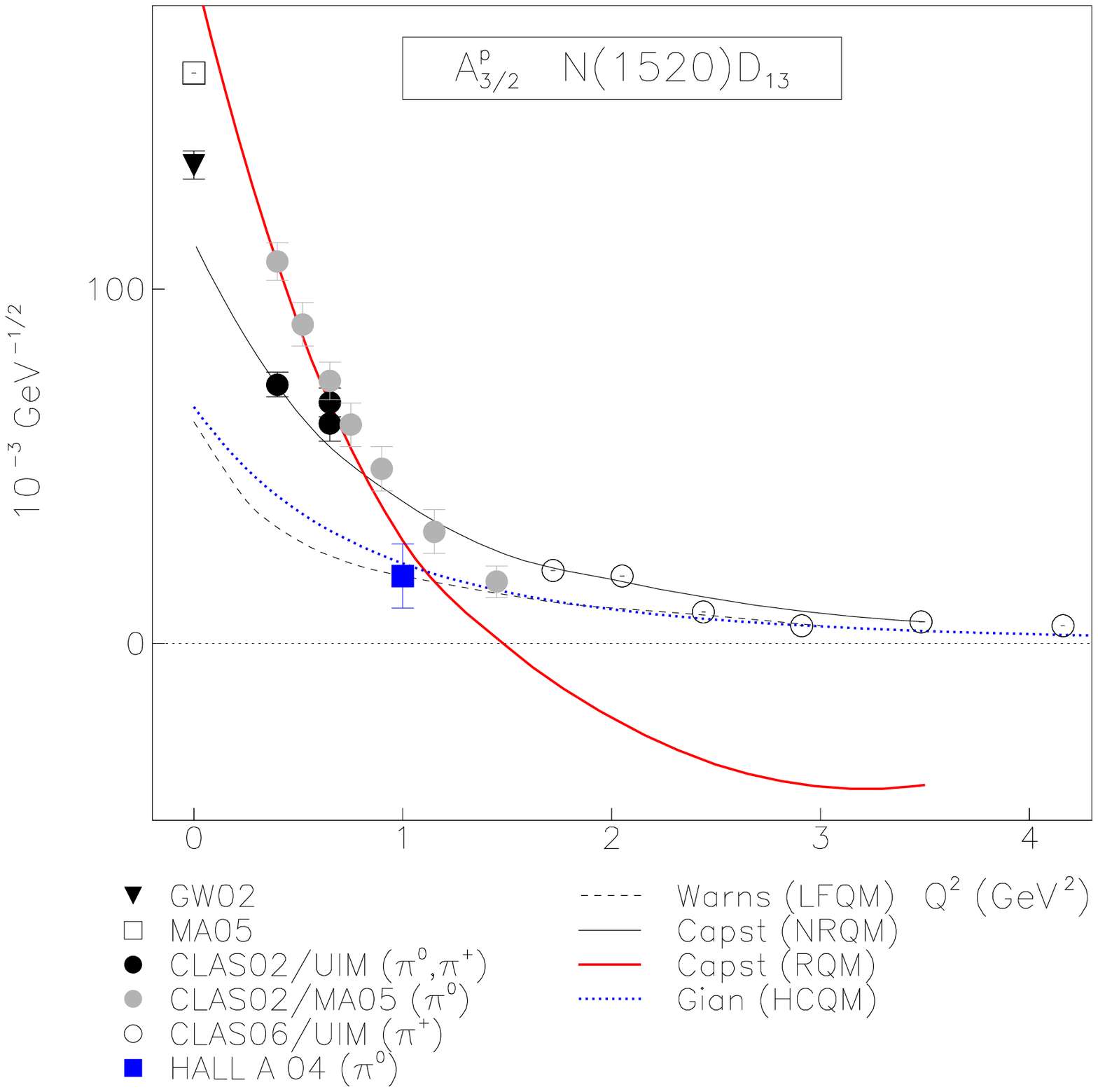}
	    \includegraphics[width=6.0cm]{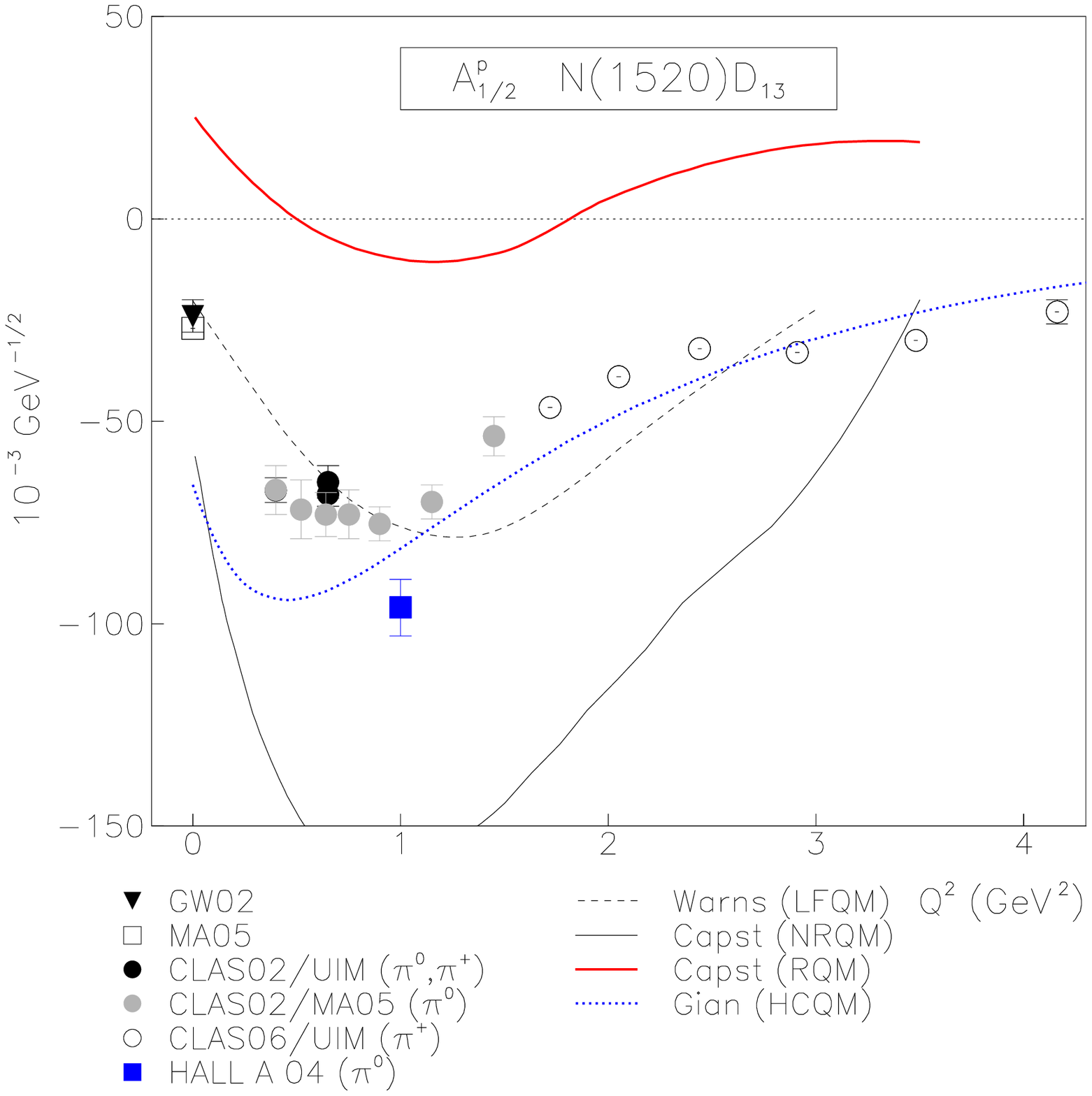}}
\caption{$Q^2$ evolution of the $\gamma p \rightarrow D_{13}(1520)$ 
transverse photocoupling amplitudes $A_{1/2}$ and $A_{3/2}$. See Figure~\ref{fig:p11}
for explantion of notation.}
\label{fig:d13}
\end{figure}

\begin{figure} 
\centerline{\includegraphics[width=6.0cm]{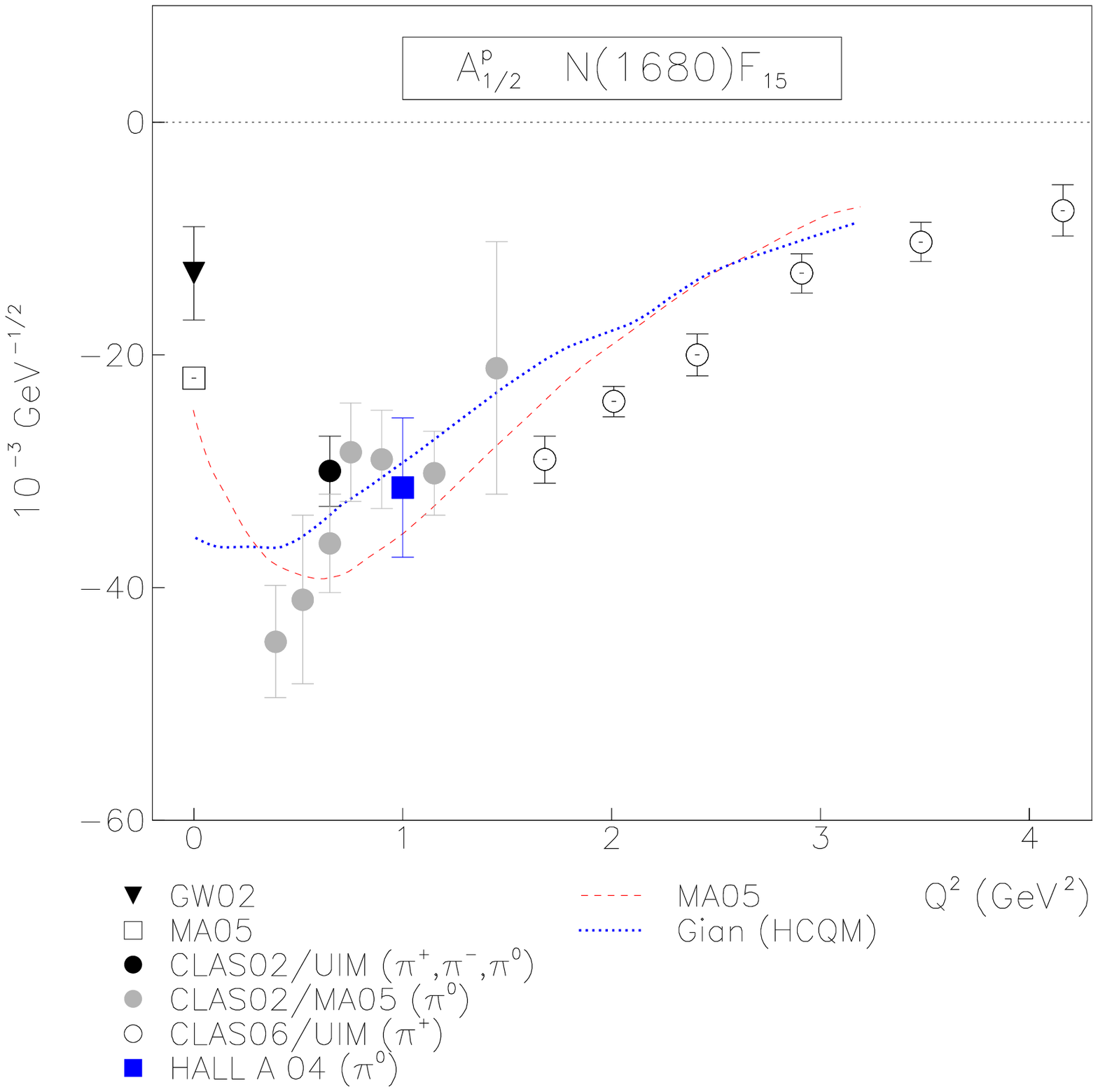}
	    \includegraphics[width=6.0cm]{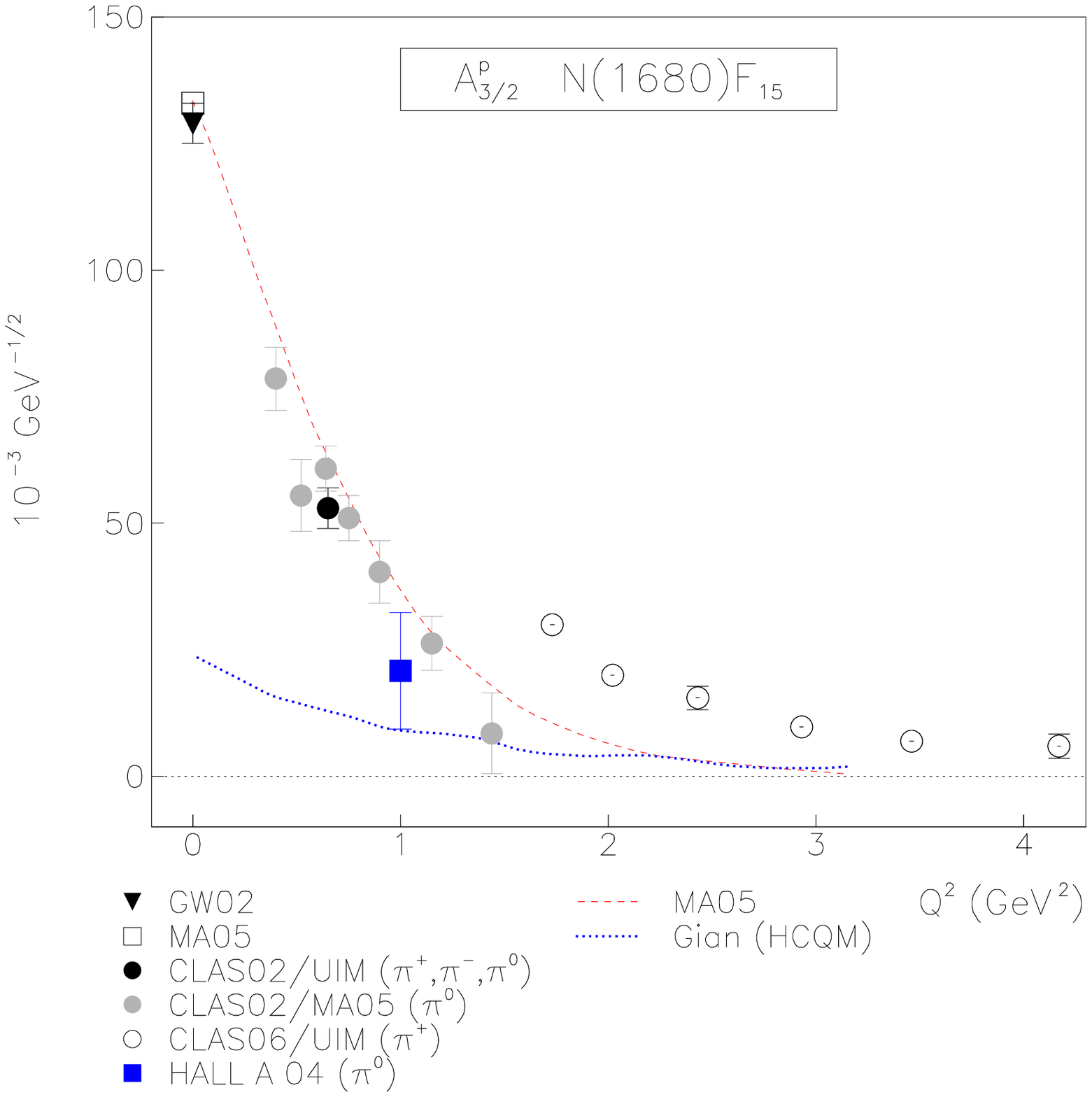}}
\caption{$Q^2$ evolution of the $\gamma p \rightarrow F_{15}(1680)$ 
transverse photocoupling amplitudes $A_{1/2}$ and $A_{3/2}$. See Figure~\ref{fig:p11}
for explantion of notation.}
\label{fig:f15}
\end{figure}

The most prominent states for $W > 1.5$~GeV are $S_{11}(1535)$, $D_{13}(1520)$, 
and $F_{15}(1680)$.  These have fairly unambiguous signatures in partial wave analyses 
and can be reasonably well isolated in single pion production channels.  
Furthermore, except for the $S_{11}$,
the $\pi N$ branching fractions and widths are known sufficiently 
to extract photocouplings with $\approx 10\%$ accuracy.  On the other hand,
a consistent quark model description of the $Q^2$ dependence of these 
three orbital excitations has proven elusive.  The new preliminary CLAS 
$p(e,e^{\prime}\pi^+)n$ data (labeled CLAS06 in Figures~\ref{fig:p11}-\ref{fig:f15}) 
has extended the study of $N\rightarrow N^*$ transition form factors into the 
unexplored region above $Q^2=2$~GeV$^2$.

For the $S_{11}(1535)$, the CLAS/UIM determination~\cite{Azn05b} of photocoupling 
amplitudes from $\pi$ and $\eta$ electroproduction channels (assuming branching
fractions $\beta_{\pi N}=0.4$, $\beta_{\eta N}=0.55$ and $\Gamma_{tot}=150$~MeV) 
give results which are in good agreement up to the highest $Q^2$ measured, unlike at 
$Q^2=0$ where $A^p_{1/2}$ from $\pi$ production is smaller than for 
$\eta$ production (see Figure~\ref{fig:s11}). It has been shown~\cite{Vra00,Arn05} 
that strong channel-coupling effects complicate the analyses near $Q^2=0$, although the
$Q^2$ evolution of these effects has not been studied.  The new data at higher $Q^2$
show a more dipole-like form factor compared to the anomalously 
hard form factor seen for $Q^2 < 2$~GeV$^2$.  Also for the first time a reliable
measurement of the longitudinal photocoupling has been made, which may shed
more light on the role of meson rescattering.

\begin{figure}	    
\centerline{\includegraphics[width=6.0cm]{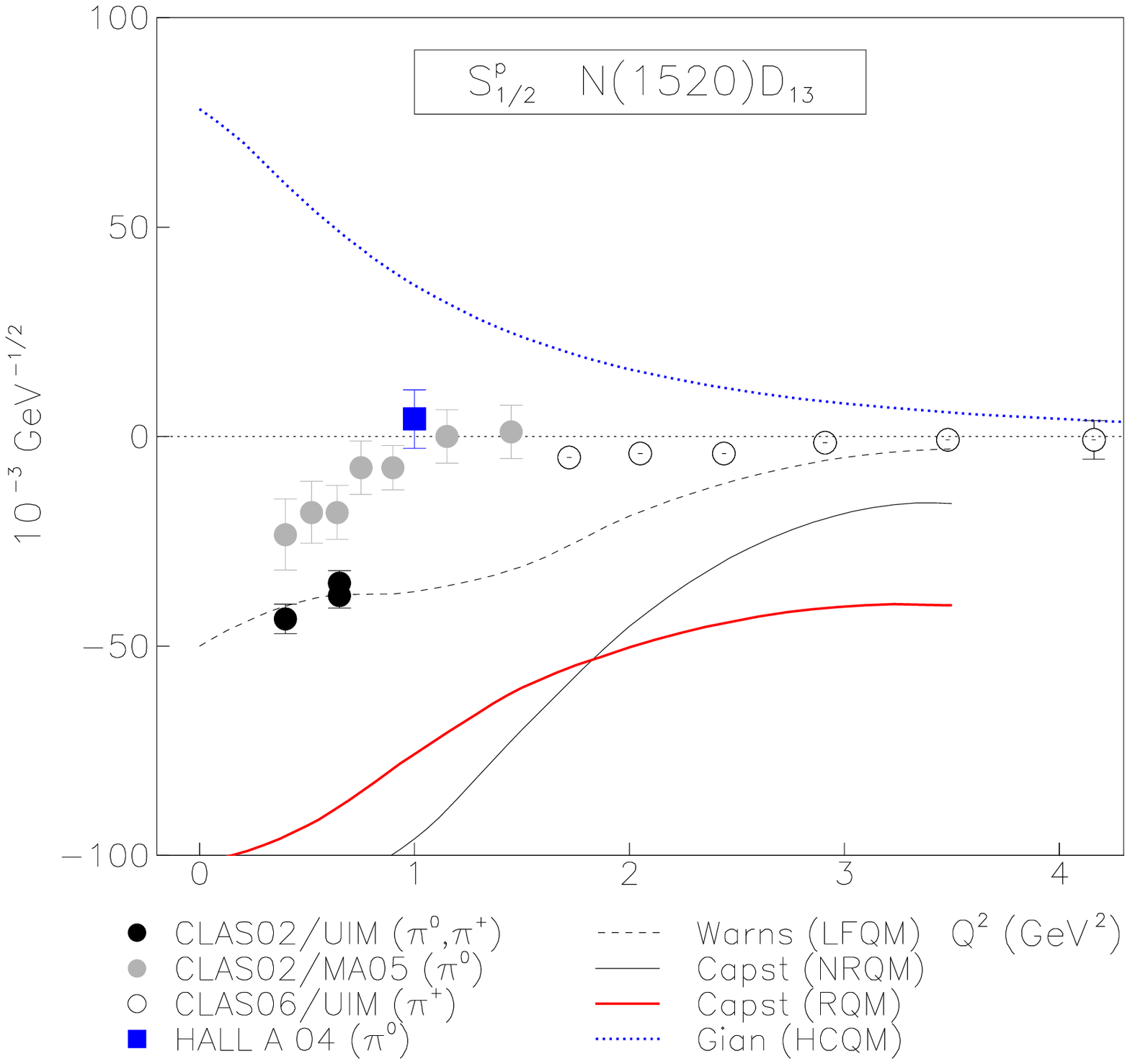}
	    \includegraphics[width=6.0cm]{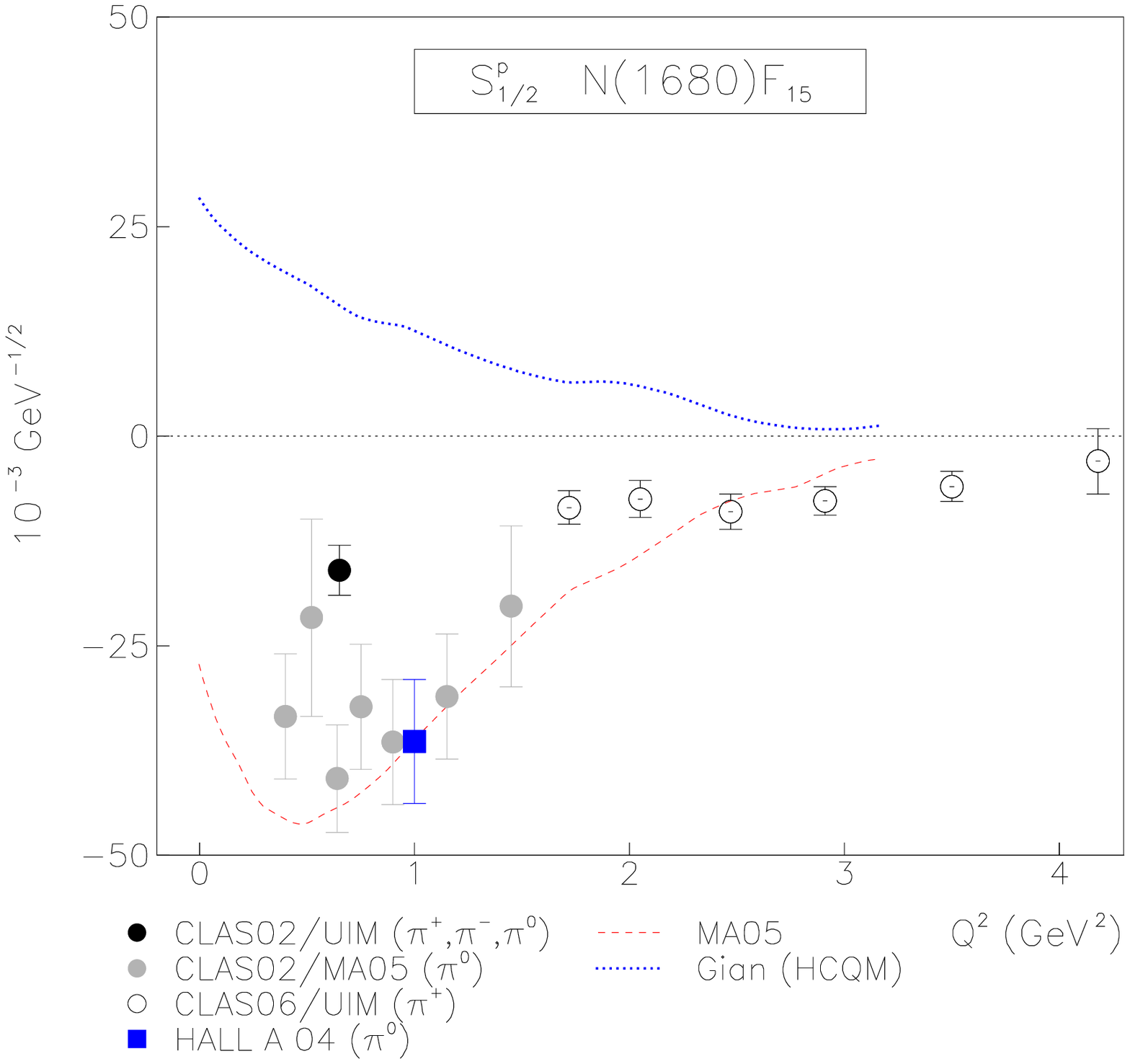}}
\caption{$Q^2$ evolution of the longitudinal photocoupling amplitudes $S_{1/2}$ 
for the $\gamma p \rightarrow D_{13}(1520)$ and  $\gamma p \rightarrow F_{15}(1680)$ 
transitions. See Figure~\ref{fig:p11} for explantion of notation.}
\label{fig:d13-f15}
\end{figure}

A more precise determination of the $\gamma^*p\rightarrow D_{13}(1520)$ and
$\gamma^*p\rightarrow F_{15}(1680)$ transitions has resulted from 
the increased $I=1/2$ sensitivity provided by the CLAS $n\pi^+$ data. 
The helicity non-conserving transverse amplitude $A^p_{3/2}$ for both 
transitions is clearly dominant at $Q^2=0$ (see Figs.~\ref{fig:d13} 
and \ref{fig:f15}), and falls very rapidly with increasing $Q^2$.  All 
quark models predict that at high $Q^2$,  the helicity conserving $A^p_{1/2}$ 
will dominate the transition.  However neither the harmonic oscillator potential
nor hypercentral potential calculations shown in the figures describe the 
$Q^2$ dependence satisfactorily.  This is especially true for the longitudinal
couplings (see also Figure~\ref{fig:d13-f15}), where the models fail to consistently
predict even the sign.  For example, quark model wavefunctions obeying
$SU(6) \otimes O(3)$ symmetry predict $S^p_{1/2}(S_{11}) \approx -S^p_{1/2}(D_{13})$~\cite{Li97}, 
in disagreement with the data.

Most higher lying states ($M>1.65$~GeV) have large $2\pi$ branching fractions 
and reliable photocoupling amplitudes can only be obtained by performing 
(at minimum) fits to measurements of both single and double pion 
production channels.  Also, detection of two pions permits the study of 
possible 'missing' resonances which have weak couplings to single meson channels. 
These broad, overlaping states in the mass region $W > 1.8$~GeV are 
predicted to decay to both $\Delta\pi$ and $N\rho$, 
which makes the $\pi^+\pi^-$ channel ideal for their detection. However, 
the reliable extraction of $N^*$ photocouplings is also hampered by a poor 
understanding of the dynamical mechanisms behind $2\pi$ production.  
Currently the most accurate data exists in the second resonance region, 
where it may be easier to disentangle the numerous resonant and non-resonant 
subprocesses which contribute to the $2\pi$ channel.

\begin{figure}[!ht]
\centerline{\includegraphics[width=7.0cm]{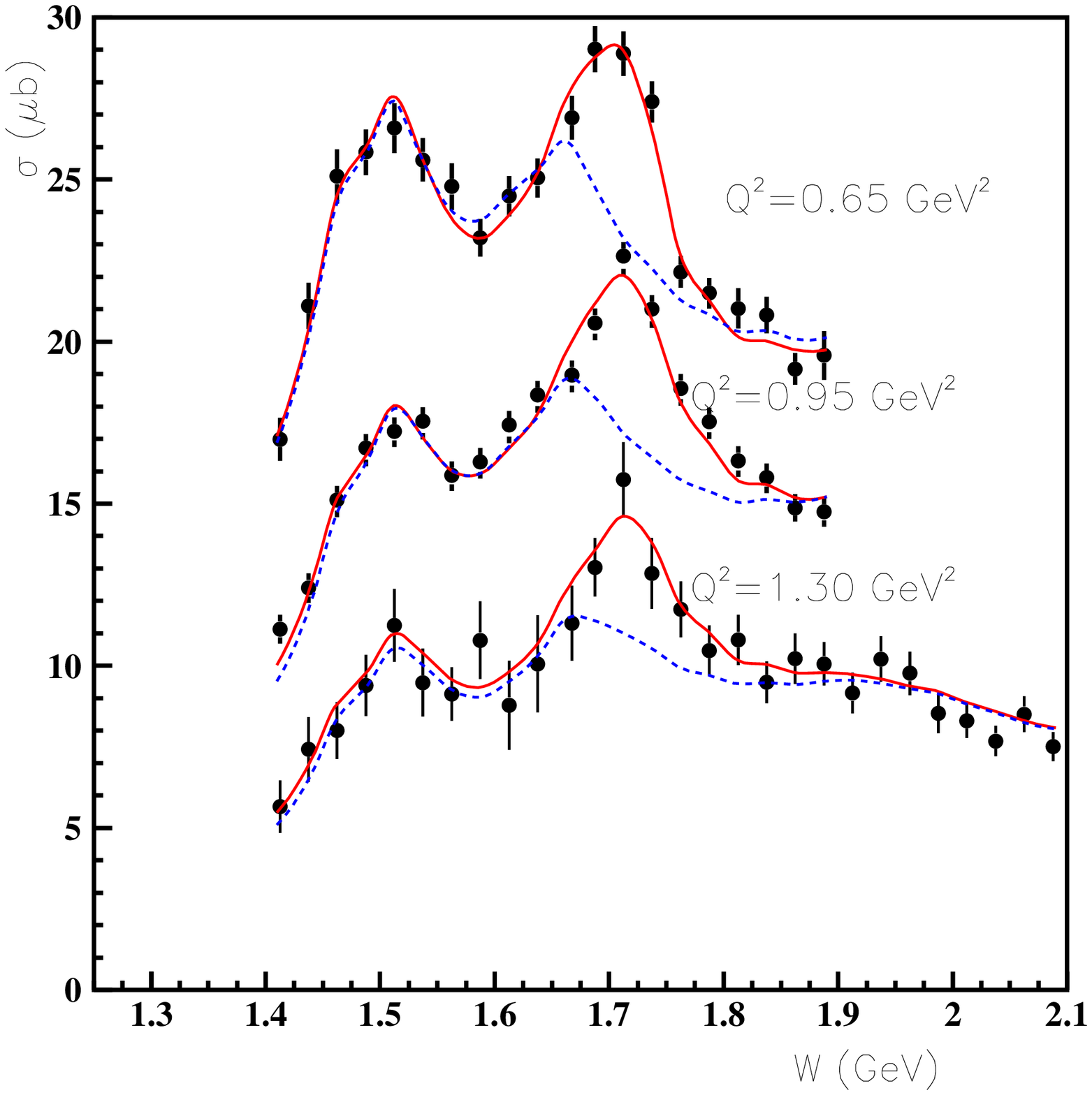}
	    \includegraphics[width=7.0cm]{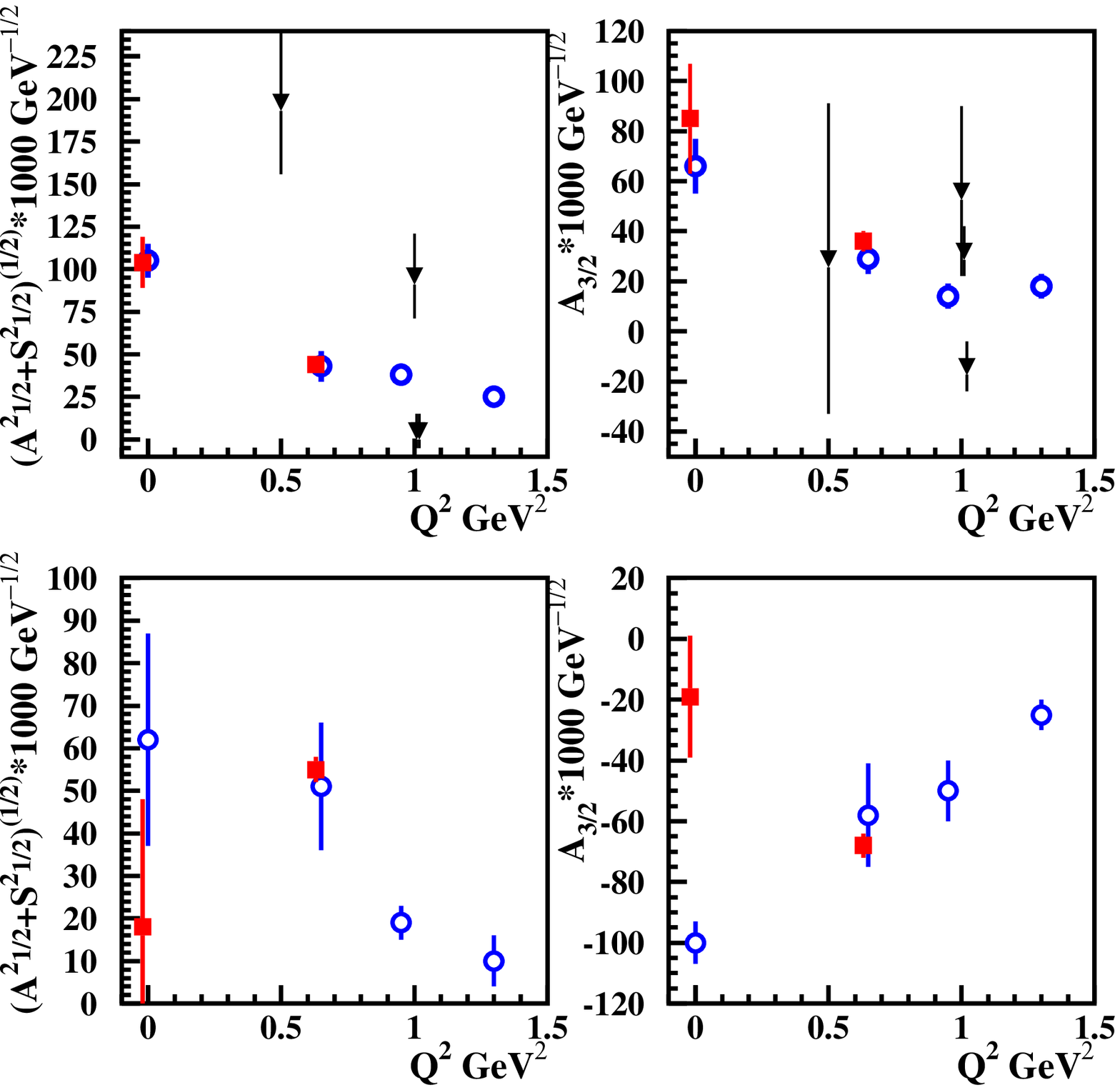}}
\caption{\small Left: Total $p\pi^+\pi^-$ electroproduction cross sections
from CLAS~\protect\cite{mokeev-1}.
Solid line: Fit using the JM05 isobar model~\protect\cite{Mo05}.
Dashed line: JM05 with new $3/2^+(1720)$ candidate state taken out.
Right: $Q^2$ dependence of photocouplings for the $D_{33}(1700)$ (top) and
$P_{13}(1720)$ (bottom) states.  Open circles are results 
extracted from CLAS $2\pi$ data~\protect\cite{Mo05}.  Other points are previous $1\pi$ analyses.}
\label{fig:twopi}
\end{figure}

The analyses~\cite{mokeev-1,mokeev-2,mokeev-3} of two pion production
by using the JLab-Moscow State University (JLAB-MSU) isobar model
considered only the minimum set of the tree diagrams proposed in
the original work of L\"{u}ke and S\"{o}ding~\cite{luke}. However, they 
made two improvements. They included all 3-star and 4-star
resonances listed by the Particle Data Group and used the absorptive model
to account for the initial and final state interactions.
They found that the $\pi N\Delta$ form factor is needed to get agreement with the data of
$\gamma p \rightarrow \pi^-\Delta^{++}$, while the initial and final
state interactions are not so large.
In analyzing the two-pion electro-production data, they further included
a $\pi\pi N$ phase-space term
with its magnitude adjusted to fit the data.
This term was later replaced by a phenomenological particle-exchange
amplitude which improves significantly the fits to the data.  Evidence
for intermediate isobar channels such as $\pi^+ D_{13}(1520)$ and
possibly $\pi^+F_{15}(1685)$ and $\pi^-P_{33}(1600)$ are suggested by
these fits.  Most prominently, for increasing $Q^2$, the $2\pi$ data show
a peak emerging at $W\approx1.7$~GeV (Figure~\ref{fig:twopi}) 
which is identified~\cite{mokeev-3} as
a $N^*$($\frac{3}{2}^+, 1720$) state.  However, this state 
differs from the conventional $P_{13}(1720)$, having
a much stronger photocoupling strength and larger $\pi\Delta/\rho N$ decay ratio
compared to PDG.  The answer to whether this is a new state or a better 
experimental determination of an existing resonance awaits further analysis.

\begin{figure}[h]
\centerline{\includegraphics[width=5.5cm]{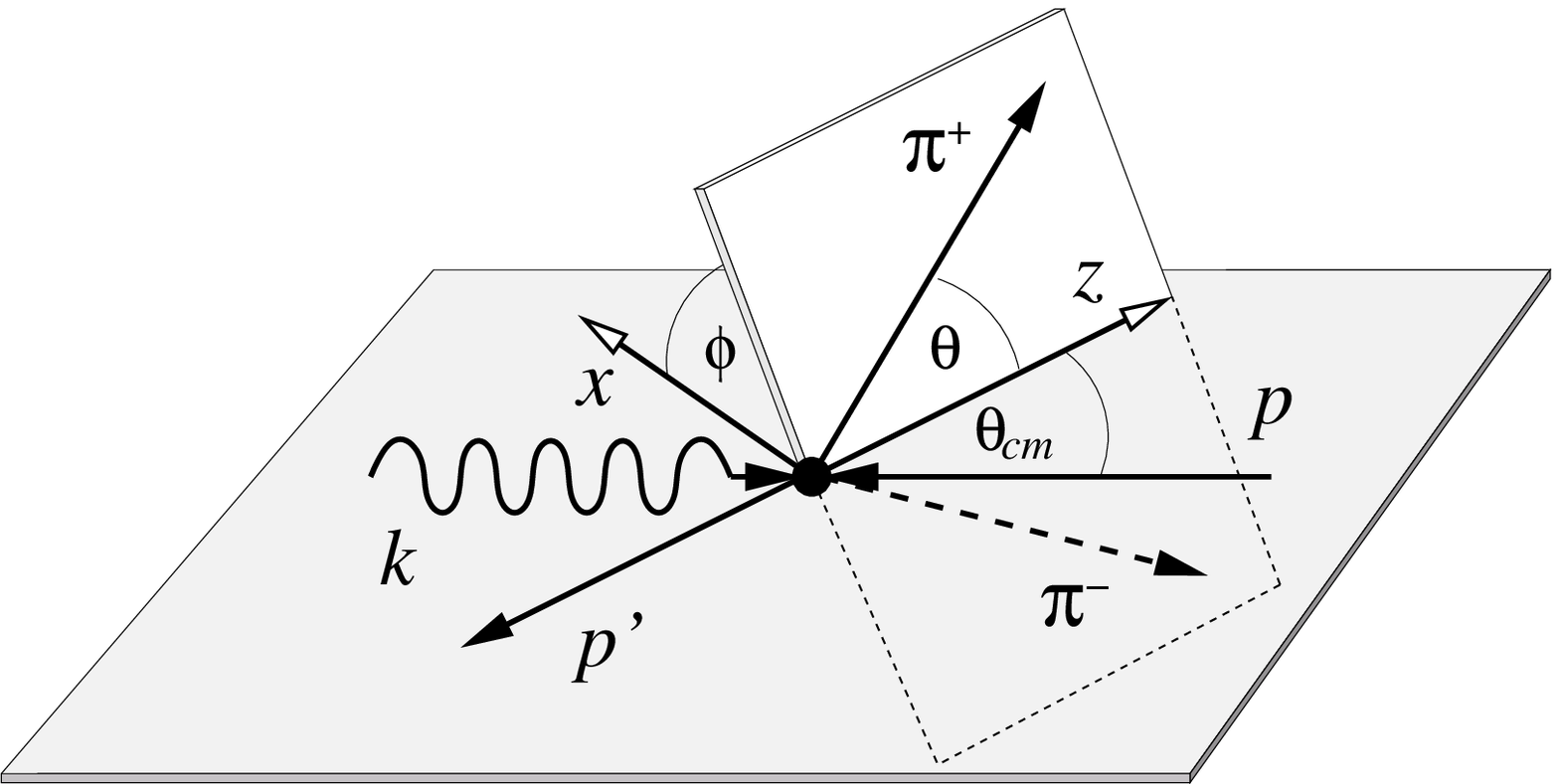}
	    \includegraphics[width=8.0cm]{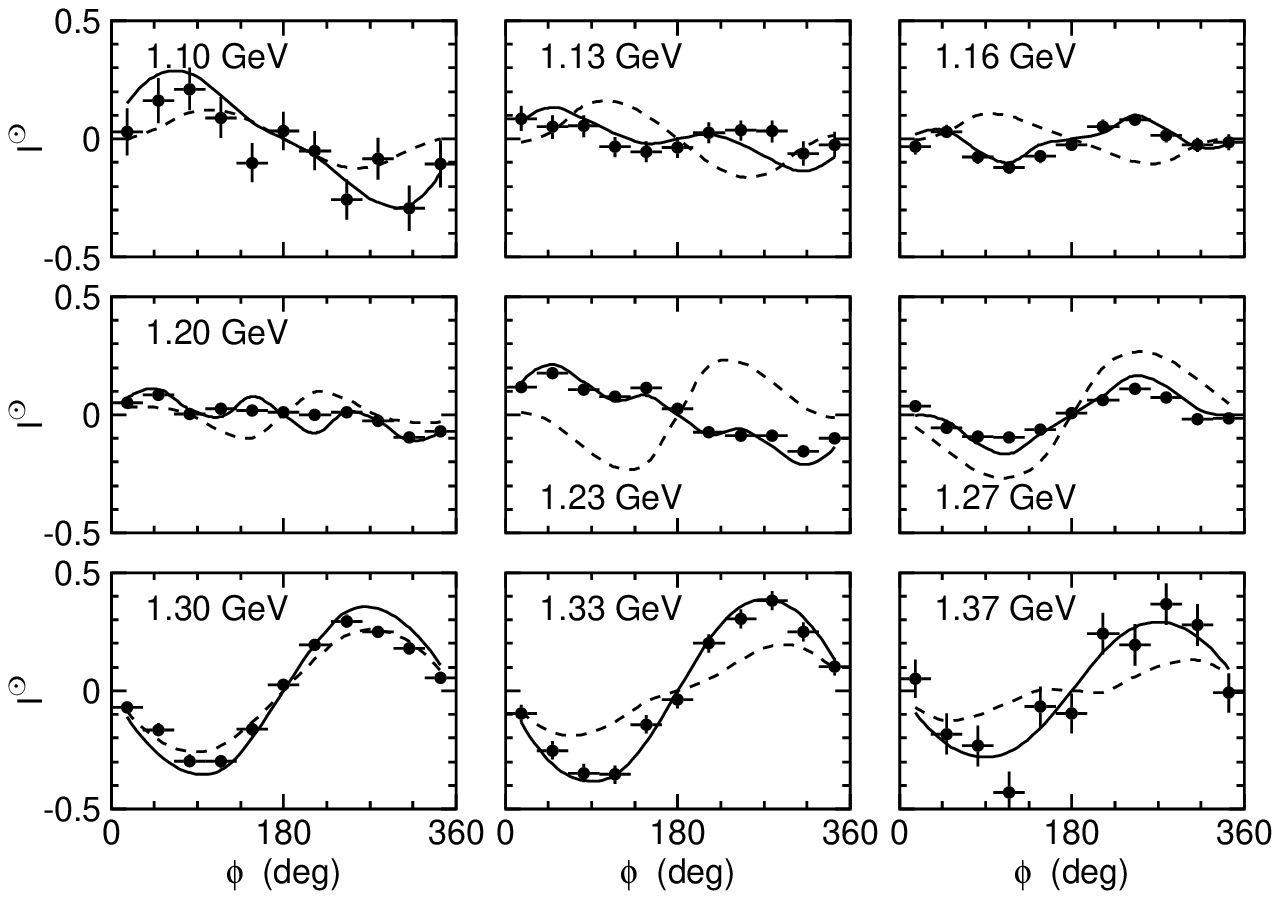}}
\caption{ Helicity asymmetries~\protect\cite{Str05} integrated over CLAS 
acceptance at $W=1.50$~GeV.  Each panel indicates the value of the invariant mass $M(p\pi^+)$. The solid, dashed
curves are calculations by Mokeev {\it et al.}\protect\cite{mokeev-2,mokeev-3} and
Fix and Arenh{\"o}vel~\protect\cite{Fix04}, respectively.  Inset shows the 
definitions of the azimuthal angle $\phi$ for the reaction 
$\vec{\gamma}p\rightarrow p\pi^+\pi^-$.}
\label{fig:2pia}
\end{figure}

\begin{figure}[h]
\centerline{\includegraphics[width=5.0cm]{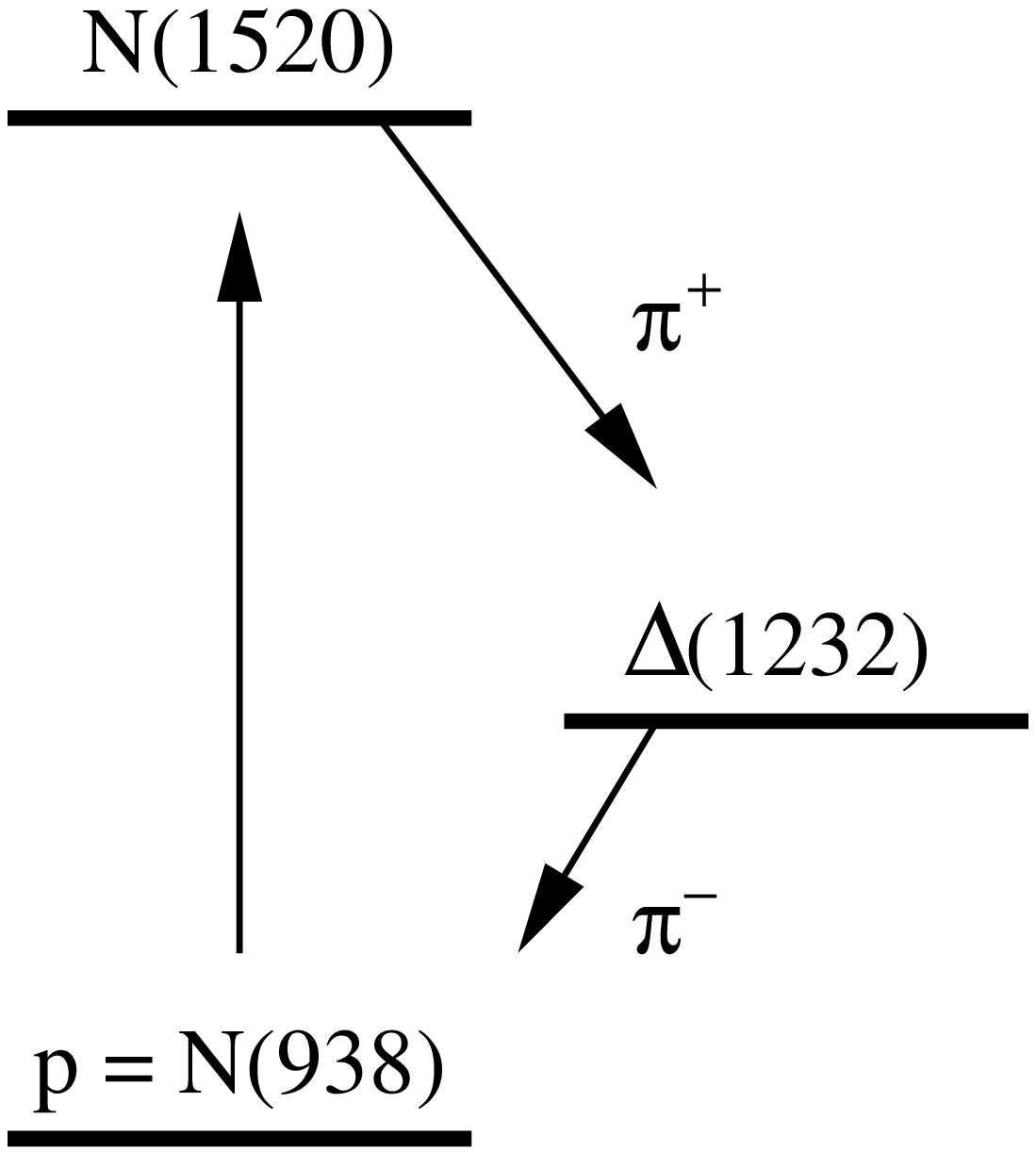}
	    \includegraphics[width=6.0cm]{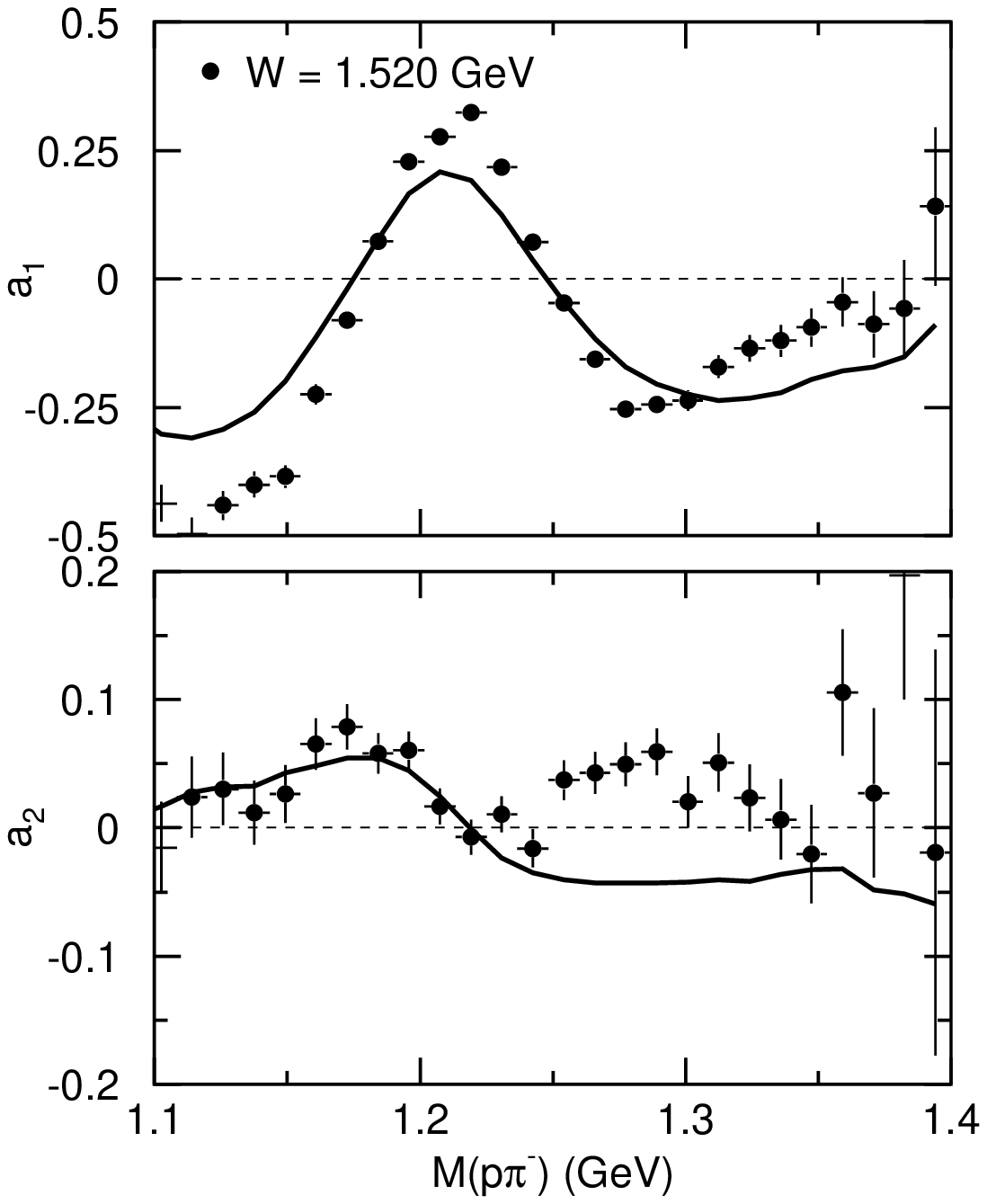}}
\caption{\label{fig:phicut}Fourier coefficients $a_1$ and $a_2$ of the helicity
  asymmetry as a function of the invariant mass $M(p\pi^-)$ for $W= 1.520$~GeV. The
  curves are the results of Fix and Arenh{\"o}vel~\protect\cite{Fix04}.
  The diagram illustrates the sequential decay $N(1520) \to \pi^+ \Delta^0 \to \pi^+\pi^-p$
  which might be observed in the data.}
\label{fig:2pib}
\end{figure}

Polarization can play an important role in the identification of 
specific subprocesses which contribute to $2\pi$ production.  
A good example is beam-helicity-asymmetry data 
in the $\vec{\gamma}p\rightarrow p\pi^+\pi^-$ reaction measured 
by Strauch {\it et al.}\cite{Str05} using CLAS with a circularly 
polarized photon beam in the mass range $W=1.35-2.30$~GeV.  
The asymmetry displayed in Figs.~\ref{fig:2pia},\ref{fig:2pib} is defined as
\begin{equation}
   I^\odot = \frac{1}{P_\gamma} \cdot 
   \frac{\sigma^+ - \sigma^-}{\sigma^+ + \sigma^-},
\end{equation}
where $P_\gamma$ is the circular polarization of the photon
and $\sigma^\pm$ are the cross sections for the photon helicity
states $\lambda_\gamma=\pm 1$.  Figure~\ref{fig:2pib} shows a Fourier
analysis of the asymmetries measured in the mass region of the $N(1520)$,
compared to model calculations incorporating $\pi\Delta$ and $\rho N$ 
quasi-two-body intermediate states.  The $\Delta(1232)$ bump seen in 
the $M(p\pi^-)$ invariant mass (Figure\ref{fig:2pib}) for 
the $a_1$ coefficient is suggestive of the
sequential decay $N(1520) \to \pi^+ \Delta^0 \to \pi^+\pi^-p$.  Previous
unpolarized cross section measurements showed strong $\Delta$ signatures 
only in the $M(p\pi^+)$ distributions, from the isospin favored 
$D_{13}\rightarrow \pi^-\Delta^{++}$ transition.  This again illustrates 
the sensitivity of polarization to the relative phases between various 
amplitudes, which can selectively amplify combinations of interfering 
background and resonance amplitudes.

\subsection{Analyses of $\omega N$, $K\Lambda$, $K\Sigma$ production data }

Because the production cross sections of the $\omega N$, $K\Lambda$, $K\Sigma$ 
channels are much weaker than those of the $\pi N$ and $2\pi N$ channels, 
tree-diagram models cannot be used reliably to extract the $N^*$ information
from the data of photo- and electro-production of these weak channels. 
We therefore review only the results from coupled-channel analyses.

Most extensive analyses~\cite{giessen3,giessen4,kvi}
 of $K\Lambda$, $K\Sigma$ and $\omega N$
photoproduction data have been  performed
by using the K-matrix coupled-channel models. 
The dynamical coupled-channel model has only been
applied to analyze~\cite{chitab,bruno} the $K\Lambda$ photoproduction.
The main focus of these analyses is to identify possible new resonances
and also to extract more accurate information on $N^*$ parameters.

The $\omega N$ photoproduction has been analyzed by the Giessen group with
the K-matrix coupled-channel model, defined by (\ref{eq:giessen}),
with $\gamma N, \pi N, \eta N,
2\pi N, \omega N$ channels. The $2\pi N$ channel is approximated
as a stable particle channel with its coupling to the $\pi N$
channel constrained by the
empirical partial-wave amplitudes~\cite{manley} of $\pi N\rightarrow \pi\pi N$.
 The interaction $V$ of (\ref{eq:giessen}) is defined
by a set of phenomenological Lagrangians with the parameters and associated
form factors adjusted to fit the empirical amplitudes
 of $\gamma N \rightarrow \pi N$, the cross sections of
$\pi N, \gamma N\rightarrow \eta N$ and 
$\pi N, \gamma N \rightarrow \omega N$.

\begin{figure}[!ht]
\vspace{6mm}
\centerline{\includegraphics[width=10.0cm]{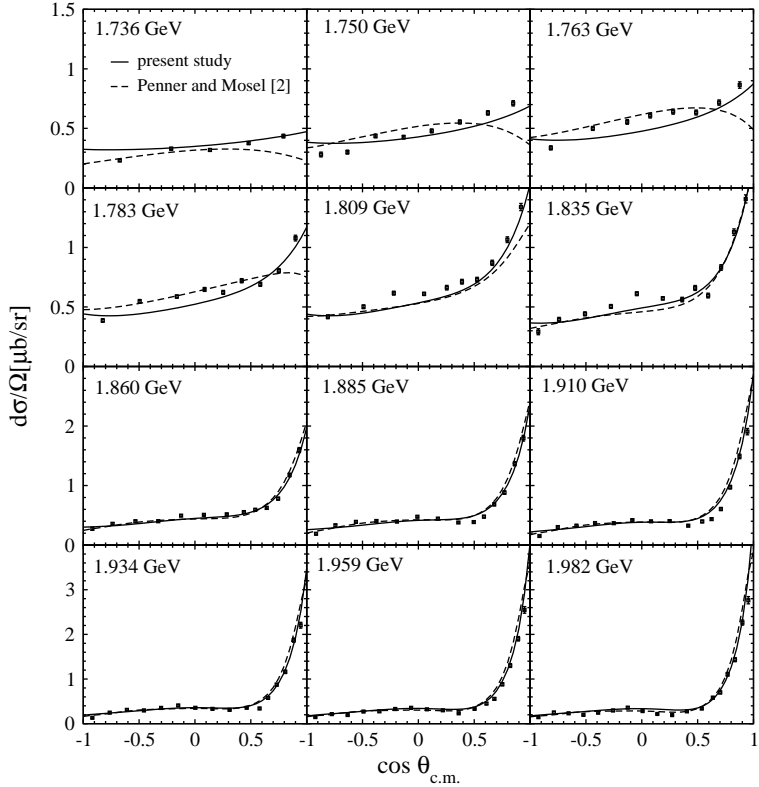}}
\caption{Fits of the coupled-channel model calculations of the Giessen 
group \protect\cite{giessen1} to experimental 
$\gamma N \rightarrow \omega N$ differential cross sections from SAPHIR~\protect\cite{bar03}.}
\label{fig:giessen}
\end{figure}

In Figure~\ref{fig:giessen}, we show their fits to the $\gamma N \rightarrow
\omega N$ differential cross sections. They found a strong contribution from
the $D_{15}(1675)$ resonance to $\pi N\rightarrow \omega N$ and from
the $F_{15}(1680)$  resonance to $\gamma N \rightarrow \omega N$.
From their analysis in Ref~\cite{giessen3}, they also extracted the
$\gamma N \rightarrow N^*$ helicity amplitudes. Their results are listed in
the first rows of Tables~\ref{table1},\ref{table2}. The SAID values from the 
VPI-GWU single-channel analysis are shown in the second rows. 
The differences between these two analysis results indicate the importance of 
including the coupled-channel effects in extracting the $N^*$ parameters.

\newcommand{\foh}{\frac{1}{2}}
\newcommand{\fot}{\frac{1}{3}}
\newcommand{\fof}{\frac{1}{4}}
\newcommand{\ftt}{\frac{2}{3}}
\newcommand{\fth}{\frac{3}{2}}
\newcommand{\ffh}{\frac{5}{2}}

\begin{table}[t]
\caption{Helicity amplitudes of $I=\foh$ resonances
( in $10^{-3}$ GeV$^{-1/2}$) determined in the Giessen study (first line);
second line: results of SAID group \cite{Arndt:1996}; "n.g.": not given.}
\begin{indented}
\item[]\begin{tabular}{l|ll|ll }
\br
$L_{2I,2S}$ & $A^{p}_{1/2}$ & $A^{n}_{1/2}$ & $A^{p}_{3/2}$ & $A^{n}_{3/2}$ \\
\mr
$S_{11}(1535)$ 
    &   92      &  -13     &  \multicolumn{2}{c}{---}  \\
    &   30(3)  &  -16(5) &   \multicolumn{2}{c}{---} \\
$S_{11}(1650)$ 
    &   57      &  -25     &  \multicolumn{2}{c}{---} \\
    &   74(1)   &  -28(4)  &  \multicolumn{2}{c}{---} \\
\hline
$P_{11}(1440)$ 
    &  -84      &  138     & \multicolumn{2}{c}{---}  \\
    &  -67(2)   &  47(5)  & \multicolumn{2}{c}{---}  \\
$P_{11}(1710)$ 
    &    -50    &   68     & \multicolumn{2}{c}{---}  \\
    &    7(15)  & -2(15)   & \multicolumn{2}{c}{---}  \\
\hline
$P_{13}(1720)$ 
     &  -65     &    1     &   35      & -4      \\
     &  -15(15) &  7(15)   &  7(10)    & -5(25)  \\
$P_{13}(1900)$ 
    &   -8      &  -19     &    0      &   6 \\
    &   n.g.      &          &           &     \\
\hline
$D_{13}(1520)$ 
    &  -13      &   -70    &  145      & -141       \\
    &  -24(2)   & -67(4)   &  135(2)   & -112(3)   \\
$D_{13}(1950)$ 
    &   11      &       40 &        26 &  -33  \\
    &   n.g.      &          &           &       \\
\hline
$D_{15}(1675)$ 
    &    9      &  -56     &    21     &  -84      \\
    & 33(4)     & -50(4)   &  9(3)    & -71(5)   \\
\hline
$F_{15}(1680)$ 
    &   3     &   30       &   116     & -48       \\
    & -13(2)  &   29(6)    &   129(2)  & -58(9)   \\
$F_{15}(2000)$ 
    &    11   &  9    &  25       &  -3   \\
    &    n.g.   &       &           &     \\
\br
\end{tabular}
\end{indented}
\label{table1} 
\end{table}

\begin{table}[h]
\caption{Extracted helicity amplitudes of $I=\fth$ resonances
( in $10^{-3}$ GeV$^{-1/2}$) considered in the Giessen study.  Second line: results of SAID group \cite{Arndt:1996}; "n.g.": not given.}
\begin{indented}
\item[]\begin{tabular}{l|l|l }
\br
$L_{2I,2S}$ & $A^{p}_{1/2}$ & $A^{p}_{3/2}$ \\
\mr
$S_{31}(1620)$ 
    &   47      &  ---   \\
    &  -13(3)   &  ---   \\
\hline
$P_{31}(1750)$ 
    &   34      &  ---   \\
    &   n.g.     &  n.g.   \\
\hline
$P_{33}(1232)$ 
    &    -128    &  -253      \\
    &    -129(1) &  -243(1)   \\
$P_{33}(1600)$ 
     &  -13     &   -28       \\
     &  -18(15) &   -25(15)   \\
$P_{33}(1920)$ 
    &    2      &    8     \\
    &   n.g.      &    n.g.    \\
\hline
$D_{33}(1700)$ 
    &  107     &  147      \\
    &   89(10) &   92(7)   \\
\hline
$F_{35}(1905)$ 
    &   32    &   -89      \\
    &   2(5)  &   -56(5)   \\
\hline
$D_{35}(1930)$ 
    &    -53   &  -23      \\
    &    4(6)  &  -3(6)    \\
\br
\end{tabular}
\end{indented}
\label{table2}
\end{table}

Experimental results for $K\Lambda$ and $K\Sigma$ production have become 
very extensive over the past decade.  However the coupled-channel analyses 
of these data are only now becoming available. The Giessen group, by extending their
model to include $KY$ channel, have found that $S_{11}(1650)$, $P_{13}(1720)$ 
and $P_{13}(1900)$ are the main resonance
contributions to $K\Lambda$ photoproduction. They show that
a peak near 1.9 GeV, which was 
identified with the "missing" $D_{13}(1960)$ in a tree-diagram model 
analysis~\cite{benn}, can be described by
a coherent sum of the contributions from the
"known" resonant and non-resonant amplitudes. 
This clearly demonstrates the importance of using a coupled-channel approach to
analyze the data of any channel, such as the $KY$ and $\omega N$ channels, having
much smaller production cross sections than the $\pi N$ and $2\pi N$ channels.

\begin{figure}[!ht]
\centerline{\includegraphics[angle=90,width=12.0cm]{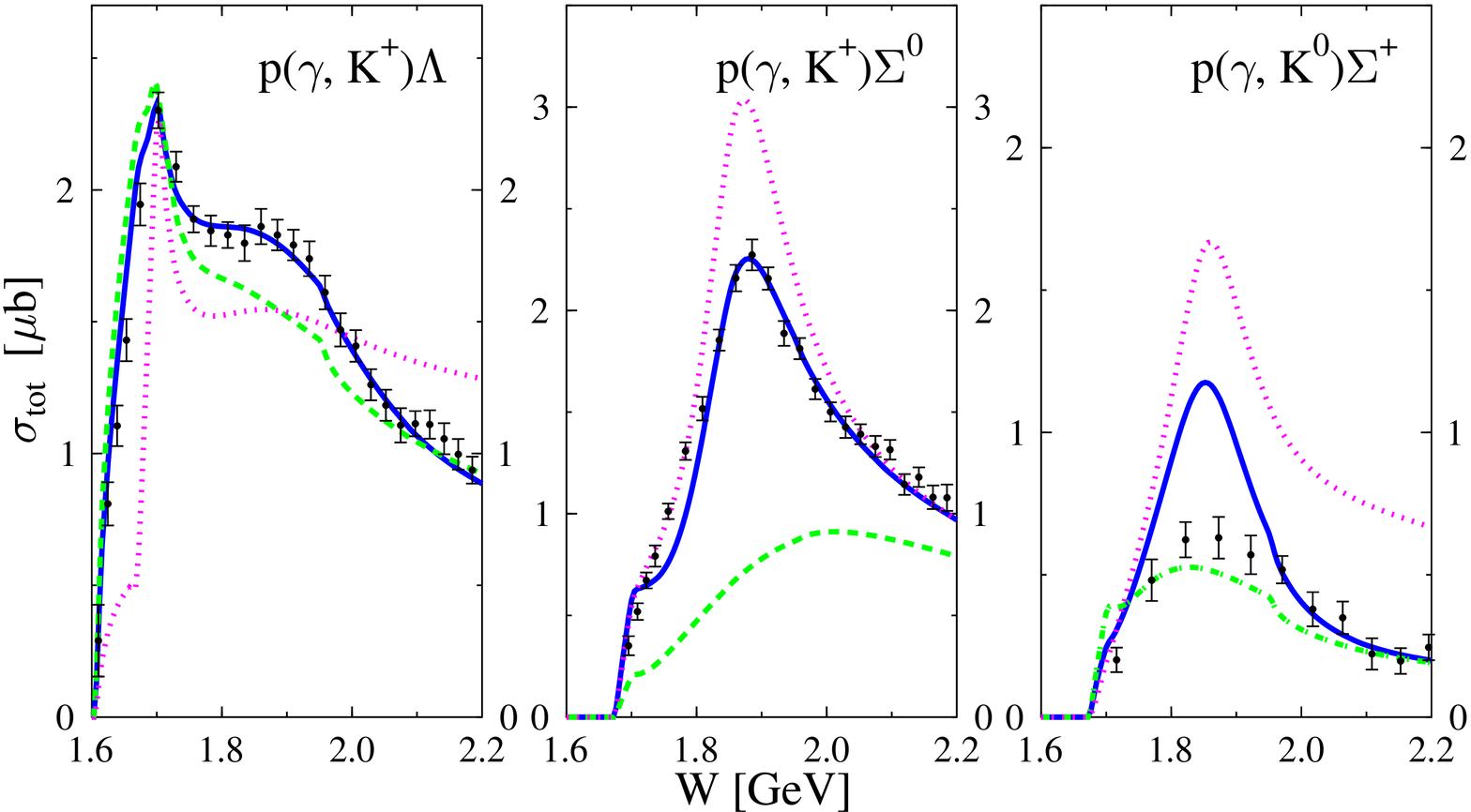}}
\caption{Experimental data on $K\Lambda$ and $K\Sigma$ total cross sections 
from SAPHIR~\protect\cite{Gla04,Law05} compared to calculations from KVI group \protect\cite{kvi}.  Dashed line shows
effect of turning off all resonance contributions.  Dotted line shows
coupled channel effects switched off.  Solid line is full calculation.}
\label{fig:klam1}
\end{figure}

\begin{figure}[!ht] 
\vspace{8mm}
\centerline{\includegraphics[width=10.0cm]{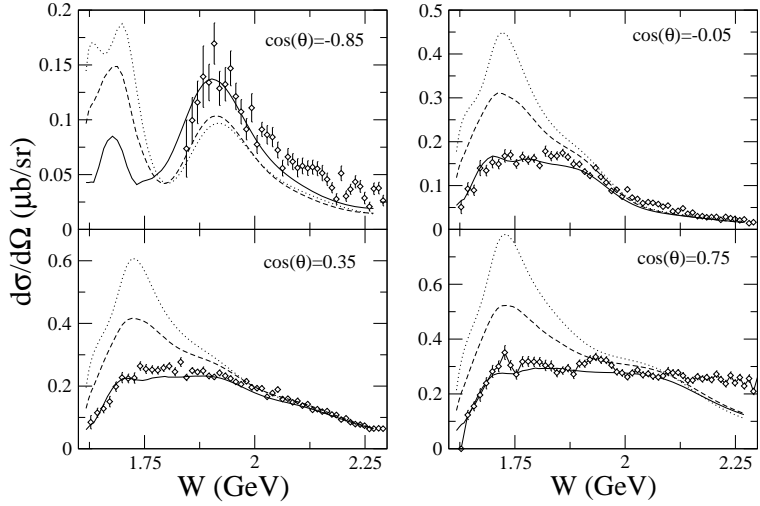}}
\caption{Comparison of the $W$ spectra of $K\Lambda$ production from CLAS~\protect\cite{Bra06} 
at various C.M. angles with model calculations from \protect\cite{bruno}.  Solid curves: full
calculation.  Dotted curves: off-shell scattering turned off.  Dashed curves: coupled-channels
turned off.}
\label{fig:jslt}
\end{figure}

The KVI group has analyzed the $K\Lambda$ and $K\Sigma$ photoproduction 
data using
a K-matrix coupled-channel model with $\gamma N$, $\pi N$, $\eta N$, $K\Lambda$,
$K\Sigma$, and $\phi N$ channels. They use
the SU(3) Lagrangian to define the interaction V of (\ref{eq:giessen}).
With appropriate choices of form factors, 
they found that the data can be described very well with
most of the coupling parameters not too different from their SU(3) limit.
Their fits to the data on $K\Lambda$ and $K\Sigma$ total cross sections
are shown in Figure~\ref{fig:klam1}. They found very large contributions from
both resonances and  coupled-channel effects. 

The $K\Lambda$ photoproduction has also been analyzed by using a
dynamical coupled-channel model~\cite{bruno} which
is based on an extension of
the dynamical model of \cite{sl} to include the $K\Lambda$
channel. It differs from the K-matrix coupled-channel models
of Giessen and KVI groups in accounting for the 
off-shell scattering dynamics.
This analysis finds that
the most relevant known resonances in $\gamma p \rightarrow K^+\Lambda$
are $S_{11}(1535)$, $P_{13}(1900)$ and $D_{13}(1520)$.  
In addition, they also demonstrate the importance of
off-shell effects and meson cloud effects in interpreting the extracted
$N^*$ parameters, as shown in Figure~\ref{fig:jslt}.

\section{Future developments}

In the past few years, we have witnessed very significant
progress in the study of $N^*$ physics.
We now have fairly extensive data for $\pi$, $\eta$, $K$, $\omega$, $\phi$ and
$\pi\pi$ production channels. Many more data will soon be available.
Theoretical
models for interpreting these new data and/or  extracting the $N^*$ parameters 
have also been developed accordingly. 

From the analyses of the single pion data in the
$\Delta$ region, 
quantitative information about the $\gamma N \rightarrow \Delta$ transition
form factors have been obtained. 
With the development of dynamical reaction models, the 
role of  pion cloud effects in determining  the $\Delta$ excitation
has been identified as the source of
 the long-standing 
discrepancy between the data and  the constituent quark model predictions.
Moreover, the $Q^2$-dependence of the $\gamma N \rightarrow \Delta$ 
form factors has also been determined up to about
$Q^2 \sim 6$ GeV$^2$.
The Magnetic $M1$, Electric $E2$ and Coulomb $C2$  $\gamma^* N \rightarrow \Delta$ transition
form factors should be considered along with the proton and neutron
form factors as benchmark data for testing various hadron models as well as
Lattice QCD calculations. 

The analyses of the $\pi$ and
$\eta$ production data had led to a rather quantitative
determination of several $N^*$ parameters in the second resonance
region. However, a correct interpretation of the $N^*$ parameters 
in terms of current hadron-model predictions requires a rigorous
investigation of the dynamical coupled-channel effects which are not 
included in amplitude analyses based on either the
K-matrix isobar model or the dispersion relation approach.

There has significant progress in analyzing
the data for $K\Lambda$, $K\Sigma$, and $\omega N$ photoproduction.
It is now generally agreed that these data must be analyzed in 
a coupled-channel approach. New information on $N^*$ parameters have
been extracted from K-matrix coupled-channel analyses. The importance of
both coupled-channel and meson cloud effects have also
been revealed. It will be interesting to extend these analyses to
analyze the electroproduction data for also extracting the $N$-$N^*$
transition form factors. 

The analysis of $2\pi N$ production data is still in a development stage, 
while some interesting results have been obtained from the analyses based on
tree-diagram isobar models.
A reliable analysis of $2\pi N$ data must satisfy the $\pi\pi N$ unitarity
condition rigorously. Progress in this direction has been made 
recently. The dynamical coupled-channel model developed in Ref.~\cite{msl}
accounts for all relevant channels and  the effects 
due to the $\pi\pi N$ unitarity condition.
It  has been applied to reveal important
dynamical features due to the $\pi\pi N$ unitary cuts which must be
included in extracting the $N^*$ information from $2\pi$ production data.
The analyses based on this dynamical coupled-channel
approach are being performed at the Excited Baryon Analysis Center (EBAC)
at Jefferson Laboratory.

Progress made in the past decade has resulted
from close collaborations between experimentalists and theorists. 
With much more complex data to be analyzed and interpreted,
such collaborations must be continued and extended in order to
bring the study of $N^*$ physics to a complete success.

\vspace{1cm}\noindent 
{\bf Acknowledgments}
This work was supported in part by the U.S. Department of Energy,
Office of  Nuclear Physics, under Contract No. W-31-109-ENG-38, and
in part under Contract No. DE-AC05-84ER40150 and 
Contract No. DE-AC05-060R23177, under which Jefferson Science 
Associates operates Jefferson Lab.

\end{document}